\documentclass[fleqn,usenatbib]{mnras}

\usepackage{newtxtext,newtxmath}

\usepackage[T1]{fontenc}
\usepackage{ae,aecompl}

\usepackage{graphicx}	
\usepackage{amsmath}	
\usepackage{breqn}
\usepackage{color}
\usepackage{cases}
\usepackage{enumerate}
\usepackage{enumitem}
\usepackage{threeparttable}
\usepackage[multidot]{grffile}
\usepackage{multirow}
\usepackage[authoryear]{natbib}
\bibpunct{(}{)}{;}{a}{}{,}

\newcommand{\BlueTides }{\textsc{BlueTides} }
\newcommand{\BlueTidesns}{\textsc{BlueTides}}

\title[The Sizes of Galaxies in the Epoch of Reionization]{The Impact of Dust on the Sizes of Galaxies in the Epoch of Reionization}

\author[M. A. Marshall et al.]{Madeline A. Marshall$^{1,2}$\thanks{E-mail: Madeline.Marshall@nrc-cnrc.gc.ca (MAM)}, Stephen Wilkins$^{3}$, Tiziana Di Matteo$^{4}$, William J. Roper$^3$, Aswin P. Vijayan$^{5,6,3}$, \newauthor Yueying Ni$^{4}$, Yu Feng$^{7}$,  Rupert A.C. Croft$^4$
\\
$^{1}$ National Research Council of Canada, Herzberg Astronomy \& Astrophysics Research Centre, 5071 West Saanich Road, Victoria, BC V9E 2E7, Canada\\
$^{2}$ ARC Centre of Excellence for All Sky Astrophysics in 3 Dimensions (ASTRO 3D), Australia\\
$^{3}$ Astronomy Centre, Department of Physics and Astronomy, University of Sussex, Brighton, BN1 9QH, UK\\
$^{4}$ McWilliams Center for Cosmology, Department of Physics, Carnegie Mellon University, Pittsburgh, PA 15213, USA \\
$^{5}$ Cosmic Dawn Center (DAWN) \\
$^{6}$ DTU-Space, Technical University of Denmark, Elektrovej 327, DK-2800 Kgs. Lyngby, Denmark \\
$^{7}$ Berkeley Center for Cosmological Physics and Department of Physics, University of California, Berkeley, CA 94720, USA \\}
\date{Accepted XXX. Received YYY; in original form ZZZ}

\pubyear{2021}

\begin{document}

\label{firstpage}
\pagerange{\pageref{firstpage}--\pageref{lastpage}}
\maketitle

\begin{abstract}
We study the sizes of galaxies in the Epoch of Reionization using a sample of ${\sim 100,000}$ galaxies from the \BlueTides cosmological hydrodynamical simulation from $z=7$ to 11. We measure the galaxy sizes from stellar mass and luminosity maps, 
defining the effective radius as the minimum radius which could enclose the pixels containing 50\% of the total mass/light in the image.
We find an inverse relationship between stellar mass and effective half-mass radius, suggesting that the most massive galaxies are more compact and dense than lower mass galaxies, which have flatter mass distributions.
We find a mildly negative relation between intrinsic far-ultraviolet luminosity and size, while we find a positive size--luminosity relation when measured from dust-attenuated images.
This suggests that dust is the predominant cause of the observed positive size--luminosity relation, with dust preferentially attenuating bright sight lines resulting in a flatter emission profile and thus larger measured effective radii.
We study the size--luminosity relation across the rest-frame ultraviolet and optical, and find that the slope decreases at longer wavelengths; this is a consequence of the relation being caused by dust, which produces less attenuation at longer wavelengths.
We find that the far-ultraviolet size--luminosity relation shows mild evolution from $z=7$ to 11, and galaxy size evolves with redshift as $R\propto(1+z)^{-m}$, where $m=0.662\pm0.009$.
Finally, we investigate the sizes of $z=7$ quasar host galaxies, and find that while the intrinsic sizes of quasar hosts are small relative to the overall galaxy sample, they have comparable sizes when measured from dust-attenuated images.
\end{abstract}


\section{Introduction}
The sizes of galaxies are a simple yet invaluable probe of the physics of galaxy formation and evolution in the early Universe. 
Galaxy sizes and morphologies are one of the small number of observational constraints with which we can currently test and refine galaxy formation models at high redshift. In addition, the assumed distribution of sizes and morphologies has an impact on the completeness of a given survey, and thus the inferred luminosity function \citep{Kawamata2018}, another key observational constraint.

In the hierarchical structure formation model, gas cools in the centres of dark matter haloes to form galaxies. 
The \citet{Mo1998} analytical model of this process predicts that if galaxies are thin exponential discs with flat rotation curves, and specific angular momentum is conserved, the scale length of a galaxy disc $R_s$ scales with redshift as $(1+z)^{-m}$ where $m=1$ at fixed halo mass, or $m=1.5$ at fixed circular velocity. Including physical prescriptions for feedback processes alters a model’s predictions for $m$ \citep{Wyithe2011}.  
At fixed redshift, the relation between galaxy size and luminosity is commonly described by
\begin{equation}
R_e=R_0 \left(\frac{L_{\textrm{UV}}}{L^*_{z=3}}\right)^\beta
\label{Eqn:SizeLum}
\end{equation}
where $R_0$ is the effective radius at $L^*_{z=3}$, $\beta$ is the slope, and
$L^\ast_{z=3}$ is the characteristic ultraviolet (UV) luminosity for $z\simeq3$ Lyman-break galaxies, which corresponds to $M_{1600}=-21.0$ mag \citep{Steidel1999}.
Analytically, \citet{Wyithe2011} predict that for a simple star formation model with no feedback, $\beta=1/3$. For models including supernova feedback $\beta$ decreases, to 1/4 for a model with supernova wind conserving momentum in its interaction with the galactic gas, or 1/5 if energy is instead conserved \citep{Wyithe2011}. 
Measuring the sizes of high-redshift galaxies thus provides valuable tests of galaxy evolution models.

A number of studies using deep Hubble Space Telescope (HST) fields have measured the sizes of $z\simeq6$--12 Lyman-break galaxies \citep[e.g.][]{Oesch2010,Mosleh2012,Grazian2012,Ono2013,Huang2013,Holwerda2015,Kawamata2015,Shibuya2015,Kawamata2018}. These find that galaxies at high redshift tend to be small, with half-light (or effective) radii ($R_e$) of typical bright galaxies being 0.5-1.0 kpc.
These observations find sizes that are consistent with an evolution of the galaxy effective radius $R_e \propto (1+z)^{-m}$ at fixed luminosity, with measurements of $m$ typically in the range of $1\lesssim m \lesssim 1.5$. \citep[e.g.][]{Bouwens2004,Oesch2010,Ono2013,Kawamata2015,Shibuya2015,Laporte2016,Kawamata2018}.
Moreover, there is growing consensus of a positive size--luminosity relationship, although with wide scatter in the $\beta$ values determined through different observations. For example,  \citet{Huang2013} found $\beta=0.22$ for $z\simeq4$ and $\beta=0.25$ for $z\simeq5$, at $z\simeq7$ \citet{Grazian2012} found $\beta=0.3$--0.5 while \citet{Holwerda2015} measured $\beta=0.24\pm0.06$, and \citet{Shibuya2015} found $\beta=0.27\pm0.01$ at $z\simeq0$--8, with no redshift evolution. 
High-redshift observational studies will soon be transformed by the James Webb Space Telescope (JWST), which will provide dramatic improvements in both sensitivity and resolution over HST. JWST will allow for more detailed morphological studies, and with its $>2\mu$m capabilities will provide the ability to robustly probe the rest-frame UV to optical morphologies deep into the reionization epoch.

The theoretical study of galaxy sizes has a long history stretching back to the earliest models of galaxy formation \citep{Fall1980}.
One method that has been used for predicting the sizes of high-redshift galaxies is semi-analytic models, which apply analytic galaxy evolution models to dark matter haloes obtained from a gravity-only simulation. \citet{Liu2016} and \citet{Marshall2019} used versions of the \textsc{Meraxes} semi-analytic model to study the sizes of high-redshift galaxies, predicting that $\beta={0.33}$ at $z=5$--10, and $m=1.98\pm0.07$ for galaxies with (0.3--1)$L^\ast_{z=3}$ and $m=2.15\pm0.05$ for galaxies with (0.12--0.3)$L^\ast_{z=3}$ \citep{Marshall2019}. However, these models evolve the sizes of galaxies analytically, and do not provide predictions for their stellar, light  or dust distributions, or detailed morphologies.
Many recent works focus instead on hydrodynamical simulations \citep[e.g.][]{Snyder2015,Furlong2017,Ma2018,Wu2020}. In this context, galaxy sizes and morphologies are driven by almost the entire ensemble of physical processes incorporated into the model. In addition to the core physical processes, sizes and morphologies will also be affected by the choice of stellar population synthesis model, alongside the model for reprocessing of stellar emission by dust and gas.

Cosmological hydrodynamical simulations allow for a detailed analysis of the sizes of galaxies in the early Universe.
Using a suite of 15 high-resolution zoom-in simulations from the Feedback in Realistic Environments (FIRE) project, \citet{Ma2018} studied the sizes and morphologies of hundreds of galaxies at $z\geq5$, with $M_\ast\simeq10^{5}-10^{9}M_\odot$. Stellar surface density and UV and B-band surface brightness images were made for each galaxy, excluding the effects of dust-attenuation.
Their high-redshift galaxies show a range of morphologies, and appear more extended when observed in the rest-frame B-band relative to the rest-frame UV, both of which are more compact than the stellar distribution. The UV morphologies are generally dominated by bright star-forming clumps, and so the B-band images are a better tracer of the overall stellar mass distribution.
In contrast, using the \textsc{SIMBA} cosmological simulations, \citet{Wu2020} predicts similar rest-frame UV and optical sizes of high-redshift galaxies. These ($25/h$Mpc)$^3$ and ($50/h$Mpc)$^3$ simulations contain $\sim5,000$ galaxies, with larger masses of up to $M_\ast\simeq10^{10.7}M_\odot$ at $z=6$, and include the effects of dust attenuation as well as a less bursty star formation model. \citet{Wu2020} found that dust attenuation makes galaxies appear larger, due to large extinction of the highly star-forming regions. 

In this paper, we investigate the sizes of high-redshift galaxies in the \BlueTides simulation. \BlueTides \citep{Feng2015} is a large cosmological hydrodynamical simulation with a volume of ($400 /h$Mpc)$^3$, containing hundreds of thousands of galaxies in the early Universe. 
With this large volume, \BlueTides allows for a statistical analysis of massive galaxies, not possible with smaller volume simulations.
The paper is outlined as follows. In Section \ref{sec:Simulation} we give an overview of the \BlueTides simulation, and detail our methodology for extracting galaxy sizes from the simulation.
In Section \ref{sec:Results} we present our results for the sizes of galaxies in the Epoch of Reionization, considering the dependence on dust and wavelength, and the redshift evolution. 
In Section \ref{sec:Quasars} we investigate the sizes of quasar host galaxies, before concluding in 
Section \ref{sec:Conclusions}.
The cosmological parameters used throughout are from the nine-year Wilkinson Microwave Anisotropy Probe \citep[WMAP;][]{Hinshaw2013}: $\Omega_M=0.2814$, $\Omega_\Lambda=0.7186$, $\Omega_b=0.0464$, $\sigma_8=0.820$, $\eta_s=0.971$ and $h=0.697$.

\section{Simulation Data and Analysis}
\label{sec:Simulation}
\subsection{\BlueTides}
The \BlueTides simulation is a large-volume cosmological hydrodynamical simulation designed to study galaxy formation and evolution in the Epoch of Reionization. \BlueTides evolves a box of volume $(400/h ~\rm{cMpc})^3$ containing $2\times 7040^{3}$ particles from $z=99$ to $z=7$ \citep{Feng2015,Ni2019}. The dark matter, gas, and star particle initial masses are $1.2 \times 10^7/h~ M_{\odot}$, $2.4 \times 10^6/h~ M_{\odot}$, and $6\times10^{5}/h~ M_\odot$ respectively, while the gravitational softening length is $\epsilon_{\rm grav} = 1.5/$h ckpc (0.24 kpc at z = 8).

A range of sub-grid physics is implemented in \BlueTides to model galaxy formation processes. The reader is referred to the original paper~\citep{Feng2015} for a full description, with a brief summary given here. 
In \BlueTidesns, gas cools via both primordial radiative cooling~\citep{Katz} and metal line cooling~\citep{Vogelsberger2014}. 
Stars form from this cool gas, based on the multi-phase star formation model originally from \citet{SH03} with modifications following~\citet{Vogelsberger2013}. \BlueTides implements the formation of molecular hydrogen and models its effects on star formation using the prescription from \citet{Krumholtz}.
Stellar feedback is included via a type-II supernova wind feedback model from \citet{Okamoto}, assuming wind speeds proportional to the local one-dimensional dark matter velocity dispersion. 
\BlueTides also includes a model of `patchy reionization' ~\citep{Battaglia}, yielding a mean reionization redshift of $z\simeq10$, and incorporating the UV background estimated by \citet{fg09}. 
The black hole growth and active galactic nuclei (AGN) feedback sub-grid model is the same as in the \textsc{MassiveBlack I \& II} simulations, originally developed in \citet{SDH2005} and \citet{Matteo2005}, with modifications consistent with \textsc{Illustris}; see \citet{DeGraf2012a} and \citet{DeGraf2015} for full details.

To extract the properties of galaxies from \BlueTidesns, we first run a friends-of-friends (FOF) algorithm \citep{Davis1985} to find haloes. As no sub-halo finder has been run on \BlueTides due to the immense size of the simulation, to extract individual galaxies we locate black holes in the simulation and assume that each lies at the centre of a galaxy. 
Black holes are seeded in all dark matter haloes when they reach a threshold mass of $M_{\rm{Halo,seed}} = 5 \times 10^{10} /h~ M_\odot$, at the location of the densest particle \citep{DeGraf2015}, and thus this is a reasonable assumption. 
The stellar mass of each galaxy is calculated as the mass contained within three times the galaxy half-mass radius $R_{0.5}$, which is calculated using particles within $R_{200}$, the radius from the black hole containing 200 times the critical stellar mass density (the critical density of the Universe multiplied by the baryon fraction and star formation efficiency of the simulation). This $3R_{0.5}$ definition provides a simple yet generally robust method to select star particles that are contained in the main galaxy component, whereas instead using a radius of $R_{200}$ or the halo $R_{200}$ often includes neighbouring galaxies.
This method locates and measures galaxies differently to sub-halo finders, which identify particles associated with galaxies based on either density peaks \citep[e.g.][]{Dolag2009} or phase-space \citep[e.g.][]{Behroozi2013,Canas2019,Roper2020}. While these methods are more complex, they do not necessarily result in measurements that are more consistent with observational galaxy selection techniques.
We note that this method is used to locate galaxies and measure their stellar mass, although for the size and luminosity measurements throughout this work we consider all particles contained in the $6\times6$ kpc image field of view (FOV) in an observation-based approach.

\subsection{Spectral Energy Distribution Modelling}
Modelling of the spectral energy distributions (SEDs) of galaxies in \BlueTidesns, including the effects of nebular emission and dust attenuation, is described in detail in \citet{Wilkins2016,Wilkins2017,Wilkins2018,Wilkins2020}. 

In brief, we associate each star particle with an intrinsic stellar SED according to its age and metallicity. 
In this work we adopt the Binary Population and Spectral Synthesis model \citep[BPASS, version 2.2.1;][]{Stanway2018} assuming a modified Salpeter initial mass function with a high-mass cut-off of $300 M_\odot$. 

We then associate each star particle with a \ion{H}{II} region (and nebular continuum and line emission) using the \textsc{Cloudy} photo-ionisation code \citep{Ferland2017}. Here we assume a metallicity $Z$ of the \ion{H}{II} region identical to that of the star particle, a hydrogen density of
100 cm$^{-3}$ \citep{Osterbrock2006}, and that no Lyman-continuum photons escape.

For star particles with ages less than 10 Myr, we assume dust contribution from a birth cloud, with optical depth
\begin{equation}
 \tau_{\rm BC}= 2 \left(\frac{Z}{Z_\odot}\right)
\left(\frac{\lambda}{5500\text{\normalfont\AA}}\right)^{\gamma},
\end{equation}
where we assume $\gamma=-1$, i.e.  $\tau_\lambda \propto \lambda^{-1}$.

We also consider dust contribution from the interstellar medium (ISM). For each star particle we calculate the line-of-sight density of metals and convert this to a dust optical depth:
\begin{equation}
 \tau_{\rm ISM}= \kappa \left(\frac{\lambda}{5500\text{\normalfont\AA}}\right)^{\gamma}  \int_{z'=0}^{z} \rho_{\rm metal}(x,y,z')dz' 
\end{equation}
where $\kappa$ and $\gamma$ are free parameters to account for the significant uncertainties associated with the production (and destruction) of dust. For the attenuation curve we assume that $\gamma=-1$, and we use $\kappa=10^{4.6}$, which is calibrated against the observed galaxy UV luminosity function at redshift $z=7$ \citep{Marshall2020,Ni2019}.
This model naturally predicts that more massive/luminous galaxies suffer higher overall attenuation \citep{Wilkins2017, Wilkins2018}. 

In Appendix \ref{App:Dust} we explore the effect of using different ISM dust attenuation laws on the measured galaxy sizes. Overall, varying the dust model makes only a small $\lesssim 0.1$ dex difference to the size--luminosity relations, and does not affect our main conclusions.

Note that throughout this work we consider only the stellar emission, and do not include the emission from AGN.

\subsection{Sample Selection}
The \BlueTides galaxy properties, such as the star formation density, stellar mass function, and UV luminosity function, have been shown to match current observational constraints at $z=7$, 8, 9 and 10 \citep{Feng2015,Waters2016,Wilkins2017,Marshall2020,Ni2019}.

\begin{figure}
\begin{center}
\includegraphics[scale=0.7]{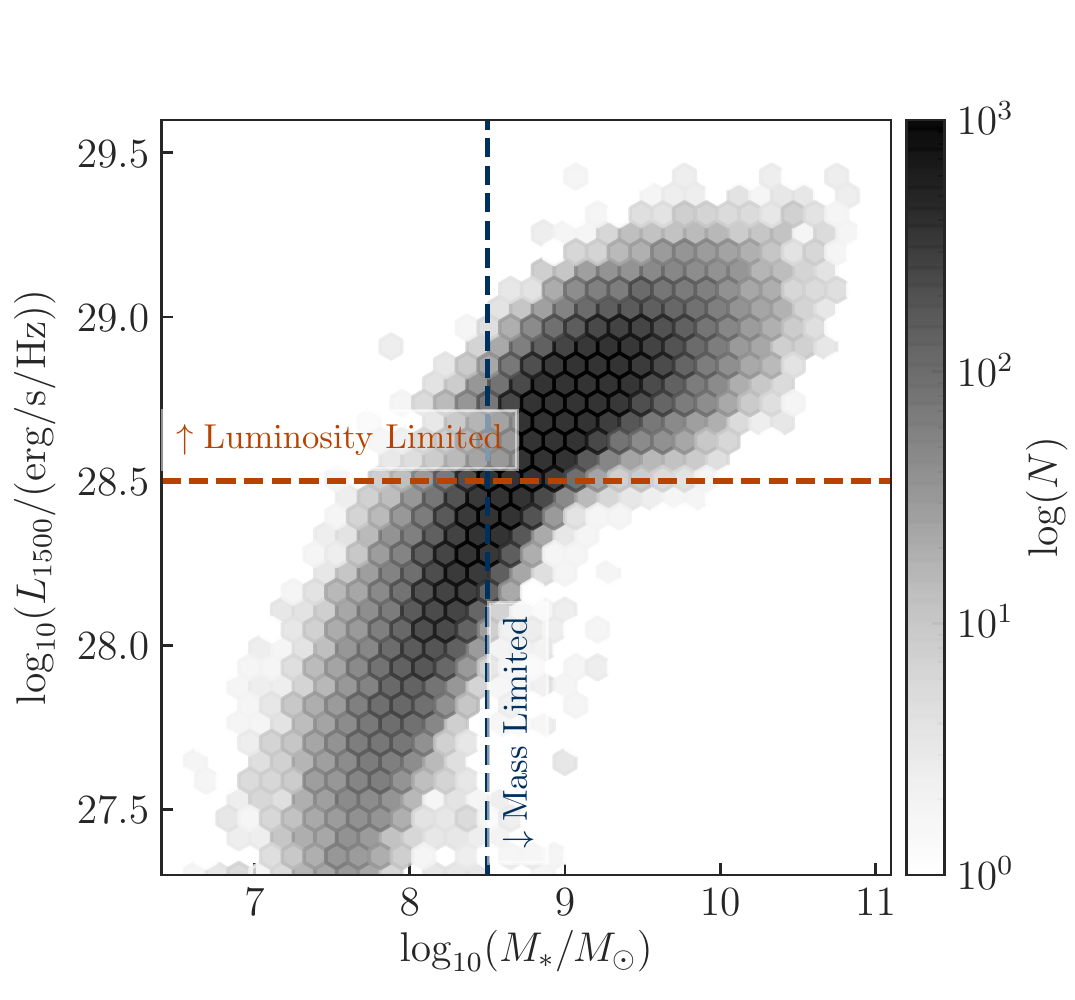}
\caption{The stellar masses and luminosities of the \BlueTides galaxy sample at $z=7$, of which there are $\sim108,000$. Also shown are the selection criteria for the two galaxy samples used throughout this work: `mass limited' galaxies with $M_\ast>10^{8.5}M_\odot$ (blue), and `luminosity limited' galaxies with $L_{1500}>10^{28.5}$ erg/s/Hz (orange). Note that the luminosity-limited sample are selected on their dust-attenuated, FUV 1500\AA~ luminosities.}
\label{fig:Samples}
\end{center}
\end{figure}

The mass--dust-attenuated FUV ($1500\AA$) luminosity distribution for the $z=7$ \BlueTides galaxies is shown in Figure \ref{fig:Samples}.
For this work we consider two samples of galaxies.
We consider a mass-limited sample of galaxies with $M_\ast>10^{8.5}M_\odot$, containing 
160 galaxies at $z=11$, 989 at $z=10$, 4,926 at $z=9$, 23,594 at $z=8$ and 79,735 at $z=7$.
We also consider a luminosity-limited sample, selecting galaxies with dust-attenuated far-UV (FUV) luminosity $L_{1500}>10^{28.5} {\textrm{erg/s/Hz}}$. This sample contains
244 galaxies at $z=11$, 1,279 at $z=10$, 5,606 at $z=9$, 22,144 at $z=8$ and 71,052 at $z=7$.
Throughout this work we consider the luminosity-limited sample when investigating luminosity trends, and the mass-limited sample when investigating mass trends.

Calculating the sizes of the large number of galaxies at $z=7$ is computationally expensive, and so for comparisons where multiple tests are run (e.g. in the Appendices), the $z=8$ sample is used.

\begin{figure*}
\begin{center}
\includegraphics[scale=0.7]{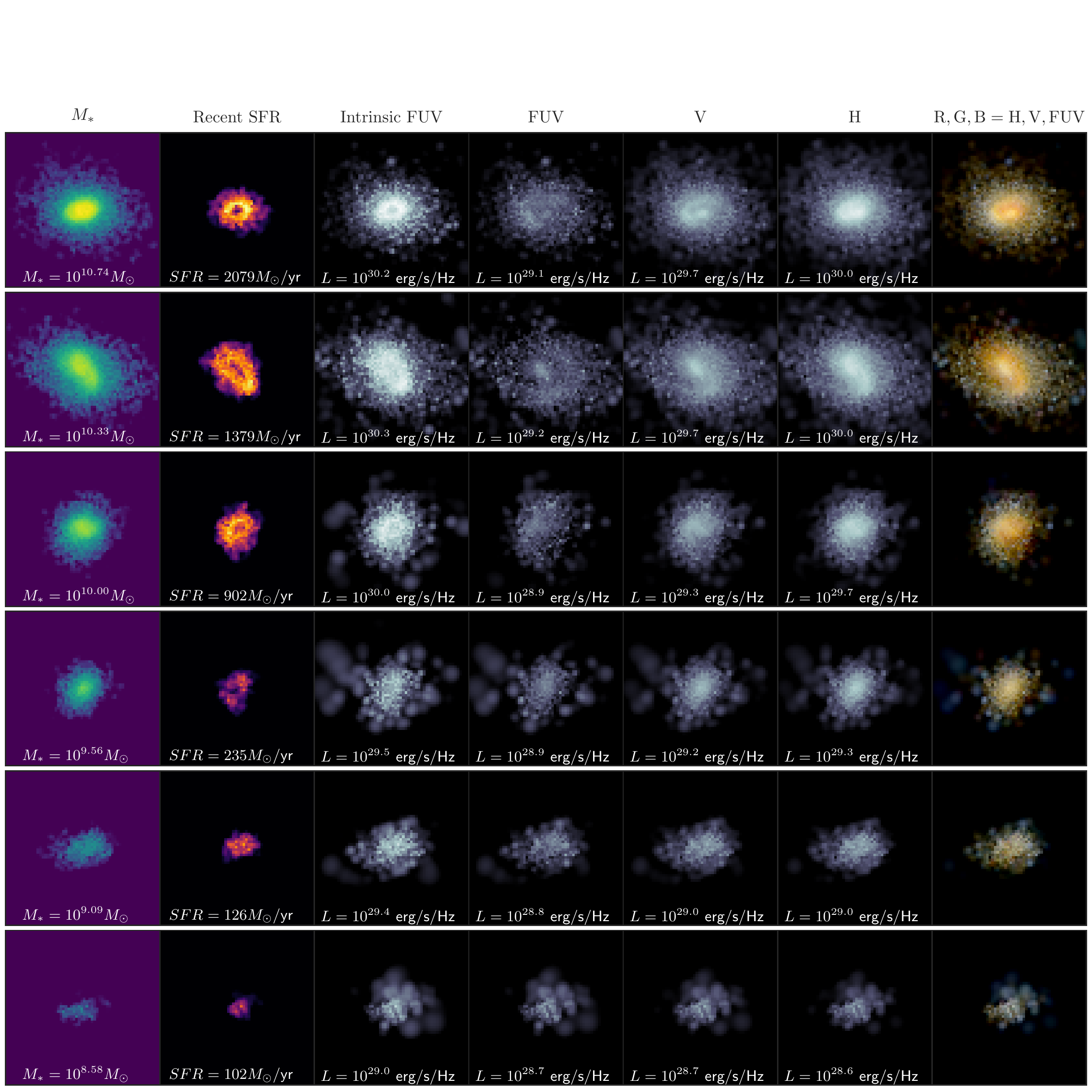}
\caption{Maps of six sample $z=8$ galaxies, ordered by stellar mass. First two columns: stellar mass and recent star formation (young stars with $<10$ Myr). Galaxy images in each of these two columns are shown on the same colour scale, logarithmically scaled from 100\% to 5\% of the brightest pixel of the most massive galaxy. 
Third column: intrinsic rest-frame FUV ($1500$\AA) images. Following 3 columns: dust-attenuated rest-frame FUV, V, and H-band images. Images in these four columns are shown on the same luminosity colour scale, a logarithmic scale with the maximum the brightest pixel in the H-band image of the most massive galaxy, and the minimum is 0.5\% of the brightest pixel in the FUV image of the most massive galaxy. 
Final column: RGB composite, from the FUV, V and H-band dust-attenuated images. The properties of each galaxy are quoted in the relevant panels. The images show a $6\times6$ kpc FOV. }
\label{fig:FullImages}
\end{center}
\end{figure*}

\subsection{Image Creation}
To create maps (images) of BlueTides galaxies we follow the approach utilised by \citet{Torrey2015} and subsequently \citet{Snyder2015}.
We smooth the light of each star particle on to a 0.1 kpc grid, adopting an adaptive approach in which the smoothing scale (full width at half maximum of the Gaussian) is equal to the distance to the 8th nearest neighbour (in 3D). 
In Appendix \ref{App:Smoothing} we also explore the assumption of a fixed smoothing scale equal to the gravitational softening length $1.5/h/(1+z)$ pkpc. We find this only makes a small difference ($\lesssim 0.1$ dex) to the overall normalisation of the luminosity--size relation and does not affect our main conclusions.

For this analysis we create rest-frame images in standard top-hat filters: FUV (1500\AA), 2500\AA, U, B, V, I, Z, Y, J and H (see e.g Figure \ref{fig:FullImages}). These images do \textit{not} include instrumental effects such as a point spread function (PSF) or noise.
We create images both with and without dust attenuation, to explore the effect of dust on the measured galaxy sizes.
We produce images with a FOV of $6\times6$ kpc. 
The images of the galaxies are made in the `face-on' direction. 
The quoted luminosity of a galaxy is the total luminosity in the corresponding $6\times6$ kpc image.
In Appendix \ref{App:FOV} and \ref{App:Orientation} we examine the effect of FOV and orientation on the measured galaxy sizes, and find that these
make only a small difference ($\lesssim 0.04$ dex) to the luminosity--size relation and do not affect our main conclusions.

Example images of six $z=8$ galaxies are given in Figure \ref{fig:FullImages}, showing their stellar mass and recent star-formation distributions, and intrinsic and dust-attenuated rest-frame FUV ($0.15\mu m$), V ($0.55\mu m$), H ($1.6\mu m$) and colour images.

\subsection{Size Measurement}

Complicating the interpretation of galaxy sizes, and specifically the comparison with observational samples, is the sensitivity of measured galaxy sizes to both the size definition and instrumental effects (e.g. noise, PSF). Common techniques used to analyse observations include curve-of-growth analyses and profile fitting either using circular or elliptical apertures (e.g. \citealt{Stetson1990}, as used in e.g. \citealt{Bouwens2004,Laporte2016}).
While useful for measuring the sizes of smoothly varying light distributions with well defined centres, these are less useful at high redshift where galaxies are generally clumpy  \citep[at least in rest-frame UV images, e.g.][]{Jiang2013b,Bowler2016}. 

In this work we adopt a non-parametric approach similar to that utilised by \citet{Ma2018}; we define an effective area $A_e$ of each galaxy as the minimum area encompassing 50\% of the total light of the galaxy in the image, even if the contributing pixels are non-contiguous. 
The effective radius of the galaxy is then defined as:  $r_e=\sqrt{A_e/\pi}$.
In Appendix \ref{App:SizeMeasurements} we compare this effective radius against the half-light radius measured from a curve-of-growth approach, and the half-mass radius measured from the intrinsic 3D stellar particle distribution from \BlueTidesns.

\section{Results}
\label{sec:Results}

\subsection{Qualitative Morphologies}
Before providing a quantitative assessment of the sizes of
galaxies in \BlueTides it is useful to explore their qualitative morphologies. 
In Figure \ref{fig:FullImages} we show example images of six $z=8$ galaxies, ordered by stellar mass, showing their stellar mass and recent star-formation distributions, and intrinsic FUV, and dust-attenuated FUV, V and H-band images. Also shown are composite colour images based on the FUV, V, and H-bands. These galaxies are chosen to span a range of stellar masses, from $\log(M_\ast/M_\odot)=$ 8.58--10.74, and they have FUV luminosities of $\log(L_{1500})=$ 28.71--29.18 erg/s/Hz.

Our least massive galaxies ($M_\ast\simeq10^{8.5}$--$10^9 M_\odot$) tend to
be clumpy irrespective of whether the map is generated from stellar mass, recent star formation, or luminosity. Conversely, our most massive galaxies ($M_\ast\simeq10^{10}$--$10^{11} M_\odot$)  generally appear smooth in the stellar mass map in addition to the longer wavelength bands (V, H) which more closely probe the underlying mass distribution. In the maps of recent star formation, galaxies are slightly more clumpy, with some examples of ring like morphology where recent star formation is confined to the outskirts of the galaxy. 
For these most massive galaxies, the dust-attenuated FUV images show more diffuse, extended emission than their intrinsic counterparts.

\begin{figure}
\begin{center}
\includegraphics[scale=0.7]{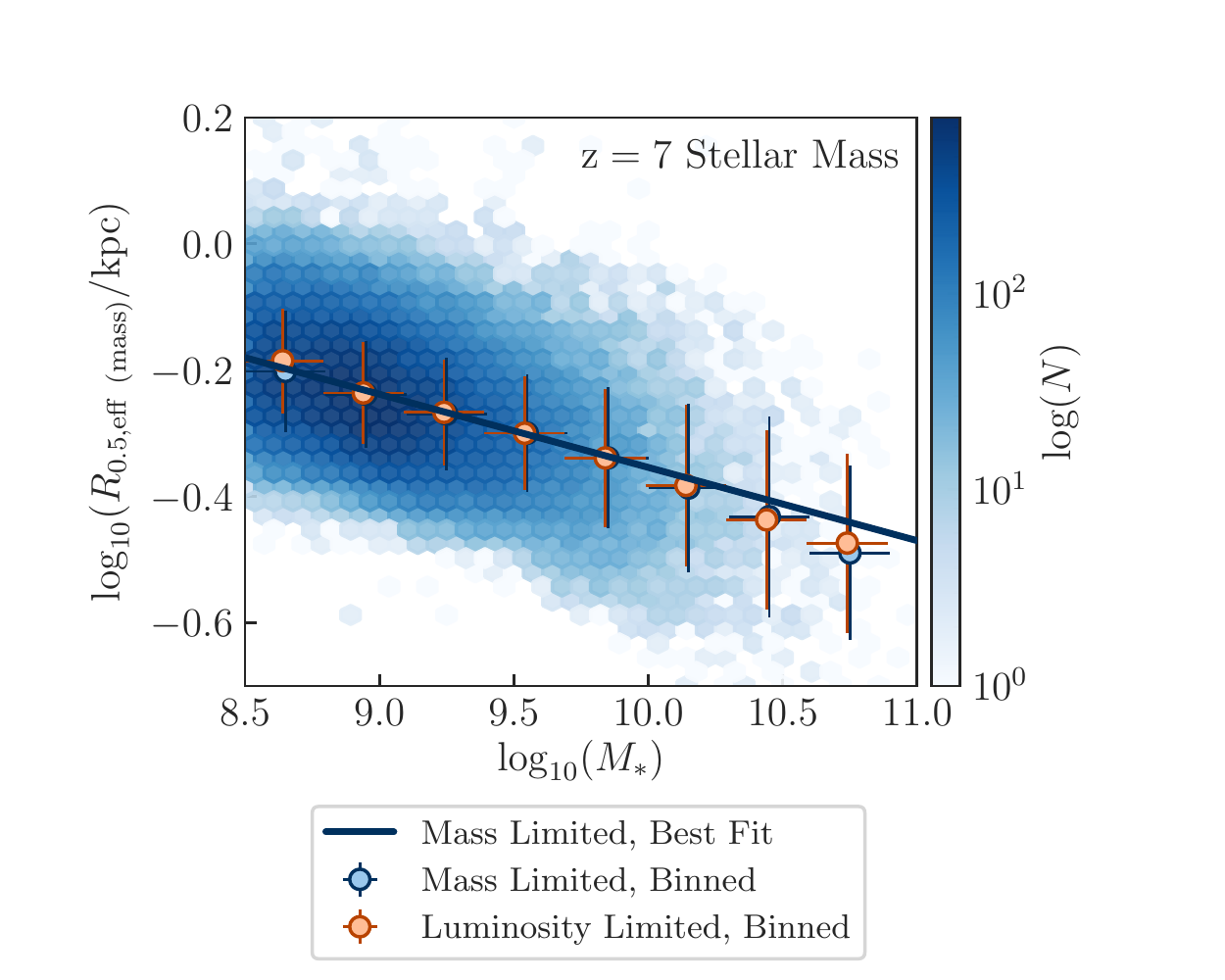}
\caption{The predicted relationship between stellar mass and size, as measured from the stellar mass map. The blue density plot depicts all galaxies with $M_\ast > 10^{8.5} M_\odot$ at $z = 7$, i.e. the mass-limited sample. The blue line shows the linear best fit to this distribution.  The blue points show the median for this mass-limited sample in bins of 0.3 dex, for bins with more than 10 galaxies, with vertical errorbars depicting the standard deviation of the distribution. The orange points show the equivalent for the luminosity-limited sample.}
\label{fig:stellarMassSize}
\end{center}
\end{figure}

\subsection{Galaxy Sizes}

\begin{figure*}
\begin{center}
\includegraphics[scale=0.7]{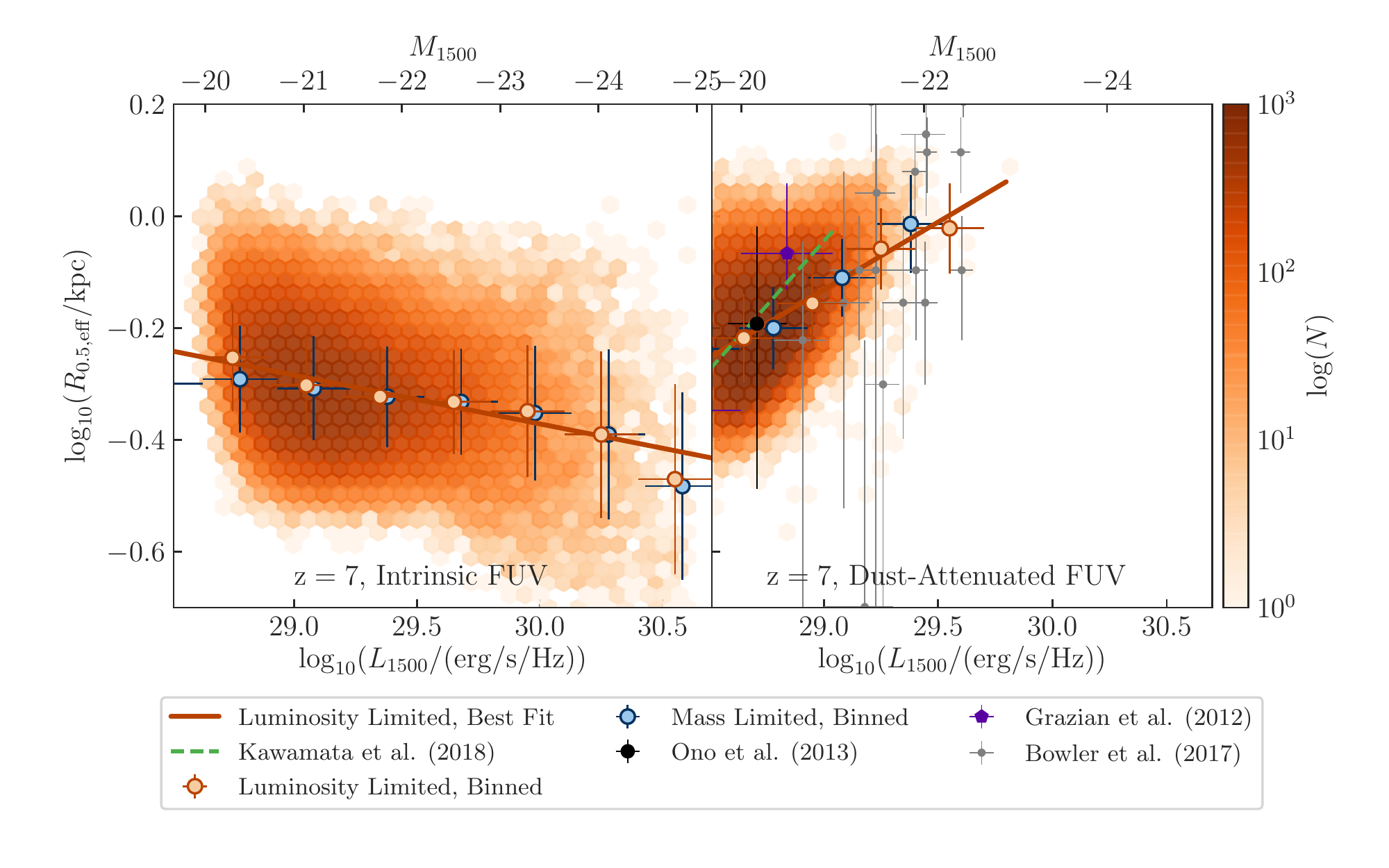}
\caption{The predicted relationship between intrinsic (left) and dust-attenuated (right) FUV luminosity ($\lambda=1500$\AA) and the measured size from that map. The orange density plot depicts all galaxies with dust-attenuated $L_{1500}>10^{28.5}$ erg/s/Hz at $z = 7$, i.e. the luminosity-limited sample. The orange line shows the linear best fit to this distribution.  The orange points show the median for the luminosity-limited sample in bins of 0.3 dex, for bins with more than 10 galaxies, with vertical errorbars depicting the standard deviation of the distribution. The blue points show the equivalent for the mass-limited sample.
The green dashed line shows the best fit to the relation observed by \citet{Kawamata2015}, and the black, purple, and grey markers depict observations of \citet{Ono2013}, \citet{Grazian2012} and \citet{Bowler2016} respectively.}
\label{fig:FUVSize}
\end{center}
\end{figure*}

\begin{figure*}
\begin{center}
\includegraphics[scale=0.7]{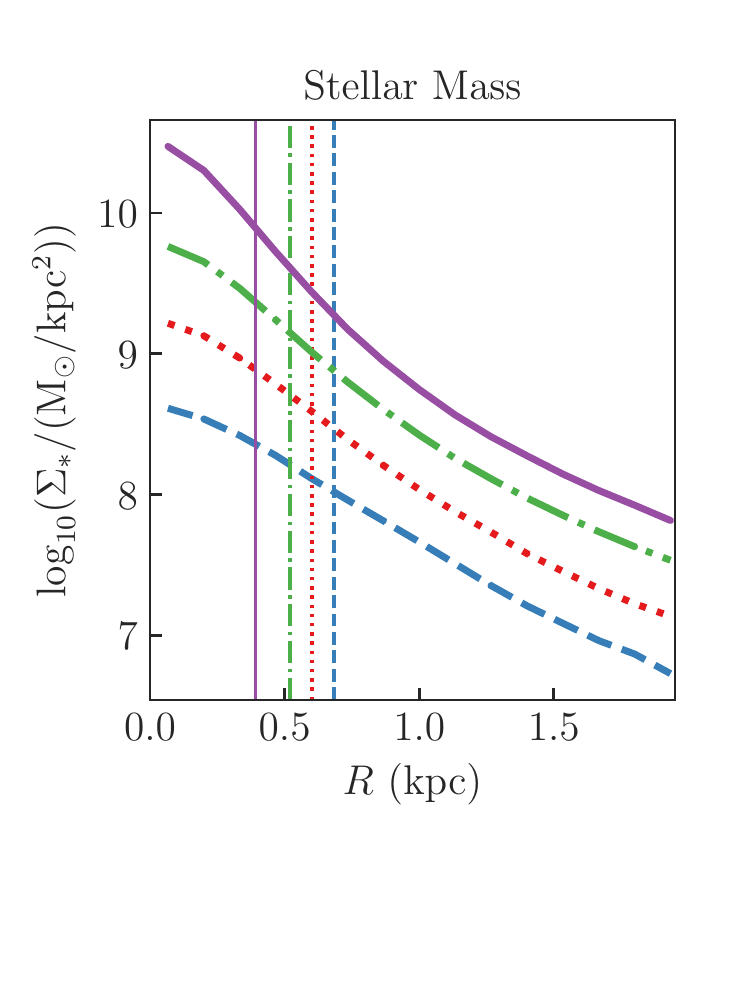}
\hspace{-0.6cm}
\includegraphics[scale=0.7]{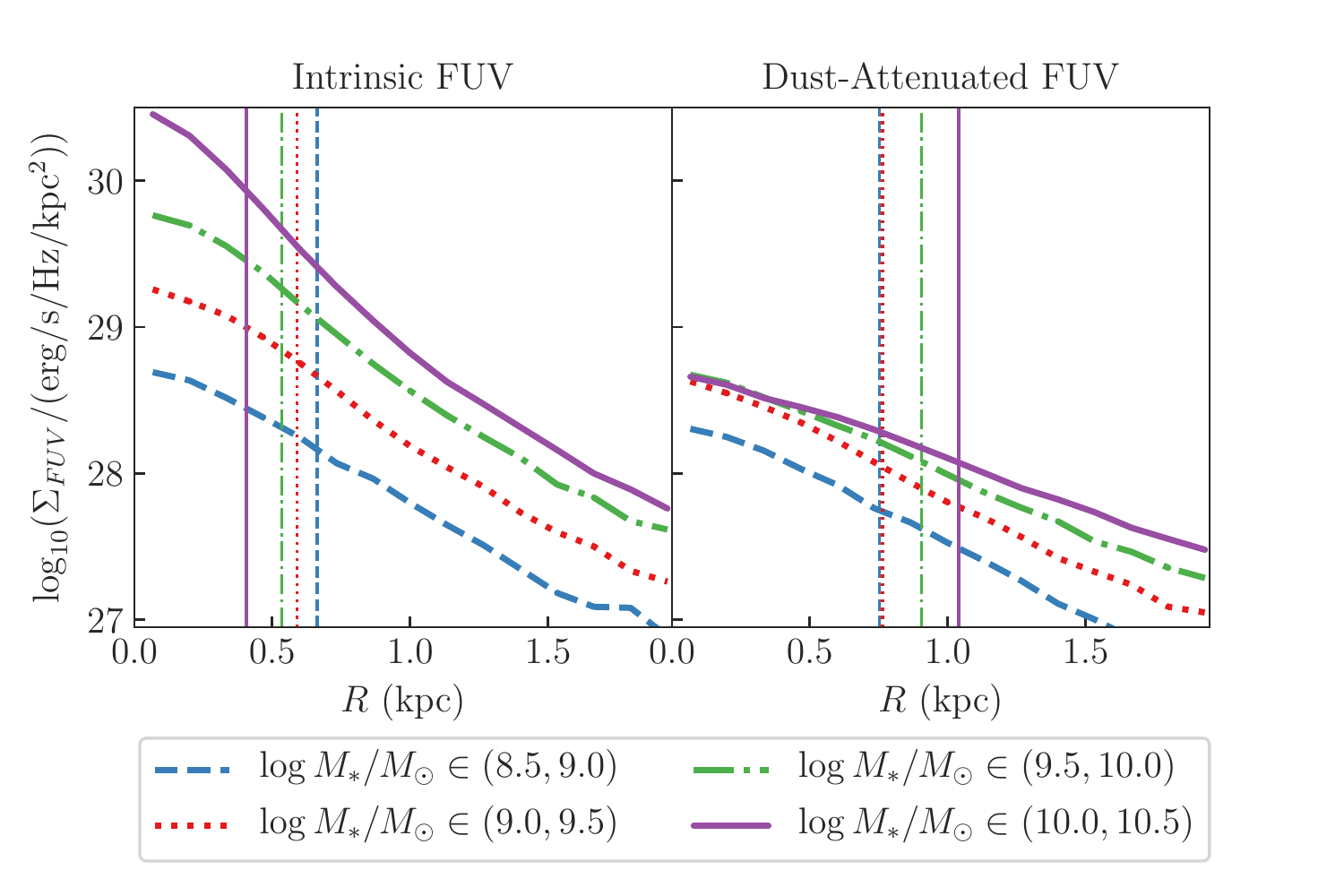}
\caption{Stacked surface density profiles of 80 randomly selected galaxies in the four mass ranges $\log(M_\ast/M_\odot)\in(8.5,9.0)$, (9.0,9.5), (9.5,10.0) and (10.0,10.5). Left: stellar mass surface density, middle: intrinsic FUV luminosity surface density, and right: dust-attenuated FUV luminosity surface density. The stacked densities are divided by the number of stacked galaxies, i.e. they represent the average surface density for the 80 galaxies.
The vertical lines depict the effective radius measured from the corresponding stacked image.}
\label{fig:surfaceDensities}
\end{center}
\end{figure*}

\subsubsection{The size--mass and size--FUV luminosity relations at $z=7$}
\label{sec:SizeMass}
We now investigate the predicted sizes of galaxies in the Epoch of Reionization. We first present results for the stellar mass--size relation predicted by \BlueTides at $z=7$, as measured from the stellar mass map, in Figure \ref{fig:stellarMassSize}. This reveals an inverse relationship between stellar mass and effective half-mass radius, suggesting that the most massive galaxies are more compact and dense.
Lower mass galaxies generally have mass distributions that are flatter and more irregular, relative to the peaked distributions of more massive galaxies (see Figure \ref{fig:FullImages}), which would result in larger measured sizes. Indeed, this is seen in the stacked surface density profiles in Figure \ref{fig:surfaceDensities}, which are discussed below.

This inverse size--mass relationship for $z=7$  \BlueTides galaxies is also seen by \citet{Marshall2020}, when considering the theoretical half-mass radius calculated from the 3D star particle distribution. 
By studying galaxies in the Illustris TNG50 simulation, \citet{Popping2021} found a slightly positive size--mass relation at $z=1$, a flat relation at $z=2$--4, and a slightly negative relation at $z=5$, consistent with our negative relation at $z=7$. For $0<z<2$ galaxies in the EAGLE simulation, \citet{Furlong2017} found that the size--mass relation is positive at $0<z<1.5$, and flattens at $1.5<z<2$, in agreement with \citet{Popping2021}.
Observations generally measure a positive size--mass relation at low-z \citep[e.g.][]{VanDerWel2014,Lange2016,Suess2019,Kawinwanichakij2021}, with weaker evidence at high-z \citep[e.g.][]{Mosleh2012,Holwerda2015}. However, some observations have found a roughly constant galaxy size with mass \citep[e.g.][]{Lang2014,Mosleh2020} which is more consistent with our findings.

Observationally it is difficult to measure total stellar masses, with the problem particularly acute at high redshift where rest-frame optical observations are only available from the Spitzer Space Telescope, which has poor resolution ($\sim2$ arcsec) compared to the typical sizes of galaxies. 
Virtually all observational size constraints are instead measured from the rest-frame FUV ($\sim1500$\AA) emission, observable with HST. 
In Figure \ref{fig:FUVSize} we show the dust-attenuated rest-frame FUV size of each galaxy as a function of its observed FUV luminosity. 
This reveals a very different relationship than for the stellar mass, with sizes increasing with the FUV luminosity.  
In Figure \ref{fig:FUVSize} we also see that this relation is in rough agreement with the observations of \citet{Ono2013}, \citet{Grazian2012}, and \citet{Bowler2016}, however the best-fitting relation from the \BlueTides predictions is shallower than that of \citet{Kawamata2018}.

To this dust-attenuated FUV size--luminosity relation we fit an equation of the form of Equation (\ref{Eqn:SizeLum}), with $R_e\propto L^{\beta}$.
We find $\beta=0.242\pm0.002$ for the luminosity-limited sample at $z=7$ (see Table \ref{table}).
This slope $\beta=0.242$ is similar to but slightly shallower than current observational estimates at $z=7$ of $\beta=0.3$--0.5 \citep{Grazian2012} and $\beta=0.24\pm0.06$ \citep{Holwerda2015}, and $\beta=0.27\pm0.01$ at $z\simeq0$--8 with no evolution \citep{Shibuya2015}. This slope is also similar to but slightly shallower than the semi-analytic predictions of \citet{Liu2016} and \citet{Marshall2019} from \textsc{Meraxes} at $z=7$, of $0.25\pm0.04$ and $0.32\pm0.01$ respectively. However, \citet{Liu2016} predict that the slope may decrease at the highest luminosities ($M_{\mathrm{UV}}\lesssim-20$ mag), corresponding to the galaxies considered in this work. \textsc{Meraxes} contains few of these luminous galaxies that are ubiquitous in the large volume \BlueTides simulation, and so comparisons are limited. 
However, it is important to note that these comparisons are problematic as we apply different size measurement and selection methodologies. 
To be completely consistent, both the simulated and observed sizes should be measured using the same methodology, including using the same detection pipelines.\\

To help understand why there is such a difference between the stellar and FUV size relations, in Figure \ref{fig:FUVSize} we show the intrinsic FUV size--luminosity relation. 
This is flatter than the relationship found for stellar mass, although still slightly negative; fainter galaxies often have smaller FUV sizes than sizes measured from the stellar mass maps, while many brighter galaxies have larger sizes in the FUV than the stellar mass maps, flattening the relation.
This may indicate varying mass-to-light ratios within and between the galaxies.
In \BlueTides the mass-to-light ratio of a star particle, which has fixed mass, is determined from the luminosity calculated using BPASS, based on the age and metallicity of the star particle. Primarily, younger stars are brighter, producing lower mass-to-light ratios, while older stars are fainter and produce higher mass-to-light ratios. 
For the lower mass galaxies shown in Figure \ref{fig:FullImages}, the bulk of the recent star formation and thus young stars are contained in the central core of the galaxy. This results in a brighter core, and a mass-to-light ratio that increases with radius, resulting in smaller measured sizes in the intrinsic FUV relative to the stellar mass map.
For the higher mass galaxies shown in Figure \ref{fig:FullImages}, the central core of the galaxy instead contains older stars, with more recent star formation occurring in rings. The stellar age distributions in these massive galaxies are consistent with the
`inside-out' galaxy formation scenario, in which the centres of galaxies assemble first and the stellar mass continues to build up around the outskirts \citep[e.g.][]{Kepner1999,vanDokkum2010,Nelson2016}, or with `inside-out' galaxy quenching, with central star formation reduced by AGN or supernova feedback \citep[see e.g.][]{Zolotov2015,Breda2020}.
This results in brighter galaxy outskirts, and a mass-to-light ratio that decreases with radius, resulting in larger measured sizes in the intrinsic FUV relative to the stellar mass map. These negative mass-to-light gradients are consistent with the \citet{Suess2019} observations of $\sim7000$ galaxies at $1\leq z\leq2.5$, which also have larger half-light than half-mass radii.
The age gradients and thus non-constant mass-to-light conversion causes both a flattening and more scatter in the size--intrinsic luminosity relation, relative to the more physical size--mass relation.

We see a much larger variation between the intrinsic and dust attenuated FUV size--luminosity relations, with the dust-attenuated relation showing a positive slope, and less scatter.
This indicates that the cause of the positive size--luminosity relationship is predominantly due to the effects of dust attenuation.
The reason for this is that in our model the dust is generally largest for bright, young stars. 
Thus light in the brightest regions of the galaxy are more heavily attenuated, resulting in a flattening of the light distribution and thus larger observed sizes when dust is included (see e.g. Figure \ref{fig:FullImages}).

To show this effect, in Figure \ref{fig:surfaceDensities} we plot the average surface density profiles of 80 galaxies randomly selected from four mass bins: $\log(M_\ast/M_\odot)\in(8.5,9.0)$, (9.0,9.5), (9.5,10.0) and (10.0,10.5). We selected 80 galaxies as there are only 83 galaxies in the largest mass range.
We see that for stellar mass and intrinsic FUV luminosity, the surface density profiles in the galaxy centres are steeper for more massive galaxies, on average. The effective sizes of these more massive galaxies are thus smaller, as half of the total mass/luminosity in the image is contained within fewer pixels. These steeper surface density profiles for more massive galaxies result in the negative size--mass and size--intrinsic luminosity relations.
Conversely, for the dust-attenuated FUV luminosity, Figure \ref{fig:surfaceDensities} shows that more massive galaxies have shallower surface density profiles.
The most massive galaxies have strong attenuation in the galaxy centre, significantly flattening the profile, whereas the least massive galaxies show less significant attenuation and have steeper dust-attenuated FUV luminosity profiles. Thus for galaxy sizes measured from the dust-attenuated FUV images, the most massive galaxies have the largest effective sizes, as half of the total luminosity in the image is contained in a larger number of pixels. Since the brightest regions of the galaxy are more heavily attenuated in our model, dust attenuation results in a flattening of the FUV luminosity surface density and thus larger observed sizes. This effect is larger for more massive galaxies, resulting in a positive size--luminosity relation.

It is important to note that we consider the effective half-mass/luminosity radius, which is a relative measure for each galaxy---this is evaluated by calculating the total mass/luminosity in the image and finding the number of pixels which contain half of this total. Hence the radius is evaluated from the shape of the mass/luminosity profile, and not its normalization. If we instead consider the extent of a galaxy as the radius beyond which the average flux drops below some fixed threshold (e.g. the noise limit of an image), we would measure larger extents for more massive galaxies. This can be visualised in Figure \ref{fig:FullImages}, with all six galaxies shown on the same mass/luminosity colour scales. The more massive galaxies have mass/flux visible out to larger radii. However, their distributions are centrally peaked while the least massive galaxies have flatter distributions, resulting in smaller measured effective radii for the more massive galaxies as discussed above. One must therefore be careful when interpreting these results---more massive galaxies on average have smaller effective half-mass radii, due to their steeper surface density profiles, resulting in an inverse size--mass relation, although the total \emph{extent} of their stellar component is generally larger than for lower mass galaxies.

In summary, while we predict an inverse size--mass relation for high-redshift \BlueTides galaxies from the stellar mass maps, the observable FUV size--luminosity relation will instead be positive due to dust attenuation.

\begin{figure}
\begin{center}
\includegraphics[scale=0.7]{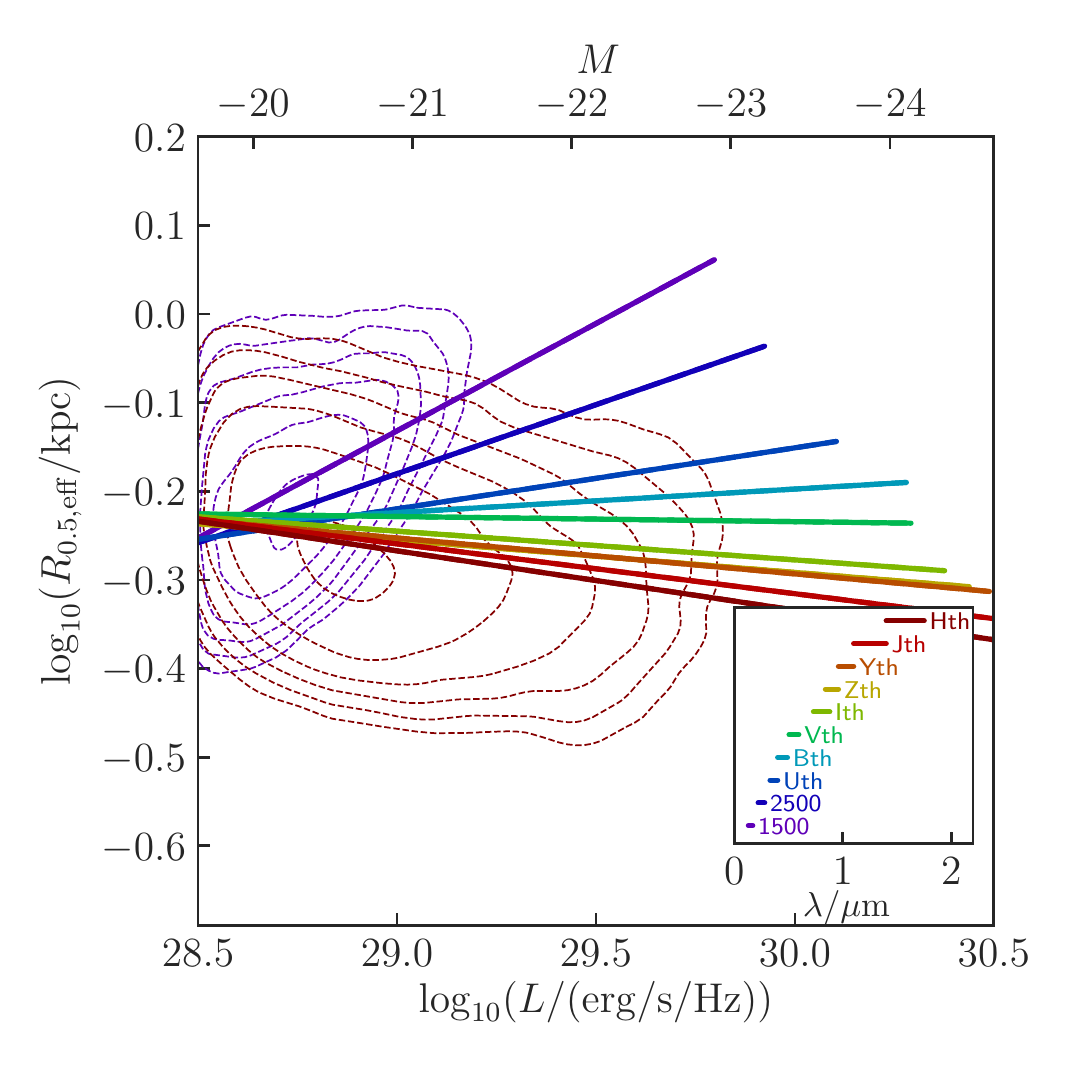}
\caption{The size--luminosity relation of the luminosity-limited $z=7$ galaxy sample for a range of rest-frame bands (with their rest-frame wavelength range depicted in the inset panel). Both the size and luminosity are measured from the image in the specified band.
Lines show the linear best fit to the relation in each band. The contours depict regions containing 
10\%, 55\%, 77\%, 88\%, 94\%, and 97\% of galaxies, for the FUV 1500\AA~ band (purple) and the H-band (red).}
\label{fig:wavelengths}
\end{center}
\end{figure}

\begin{figure*}
\begin{center}
\includegraphics[scale=0.7]{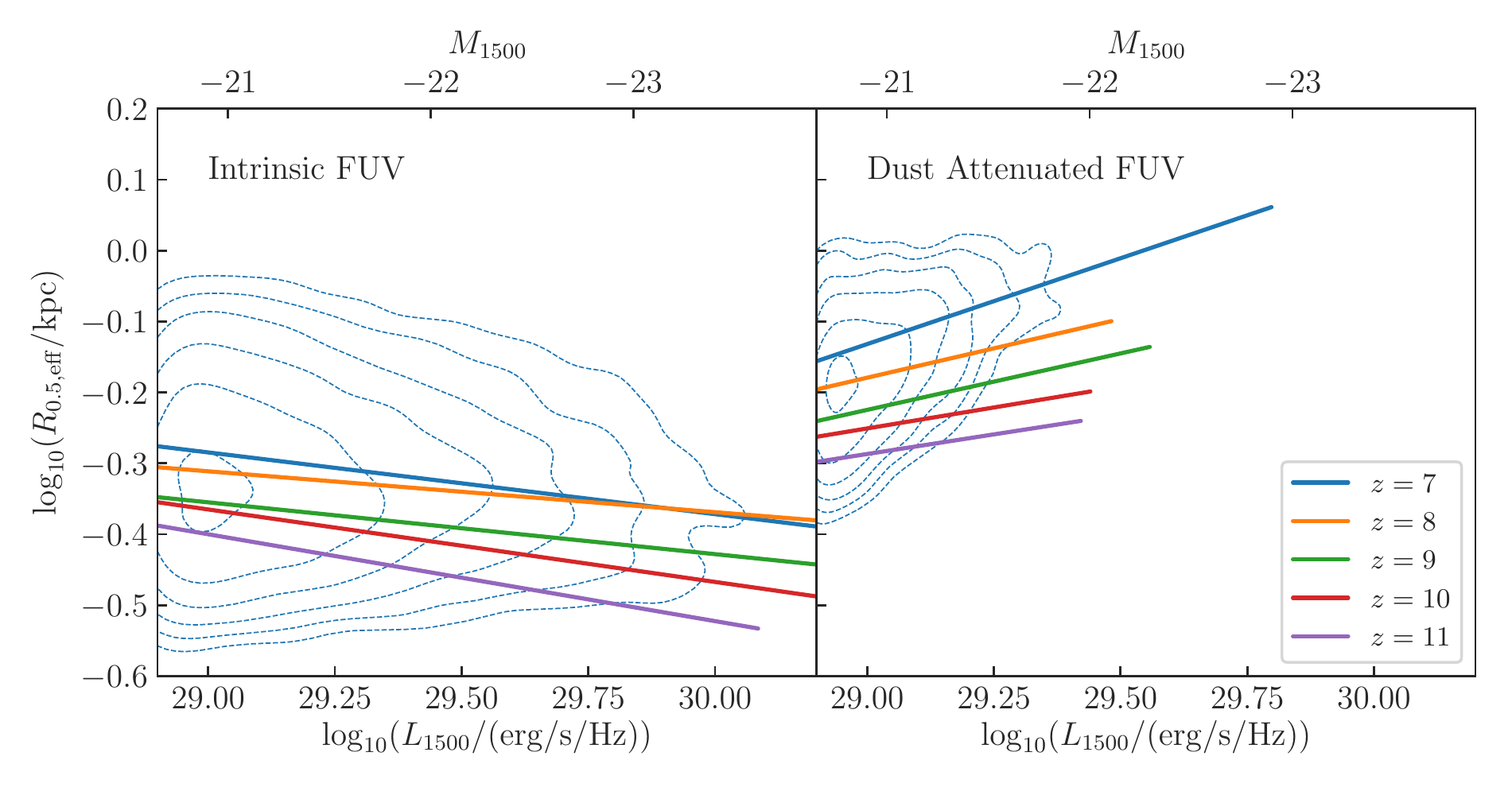}
\caption{The FUV 1500\AA~ size--luminosity relation for galaxies from $z=7$ to $z=11$, for Left: their intrinsic FUV maps and Right: the dust-attenuated maps. Galaxies are from the  luminosity-limited samples, i.e. with dust-attenuated $L_{1500}>10^{28.5}$ erg/s/Hz.
Lines show the linear best fit to the relation at each redshift. The contours depict regions containing 
10\%, 55\%, 77\%, 88\%, 94\%, and 97\% of galaxies for $z=7$. The luminosity to magnitude conversion corresponds to $z=7$.}
\label{LuminositySizeRedshift}
\end{center}
\end{figure*}

\begin{figure}
\begin{center}
\includegraphics[scale=0.7]{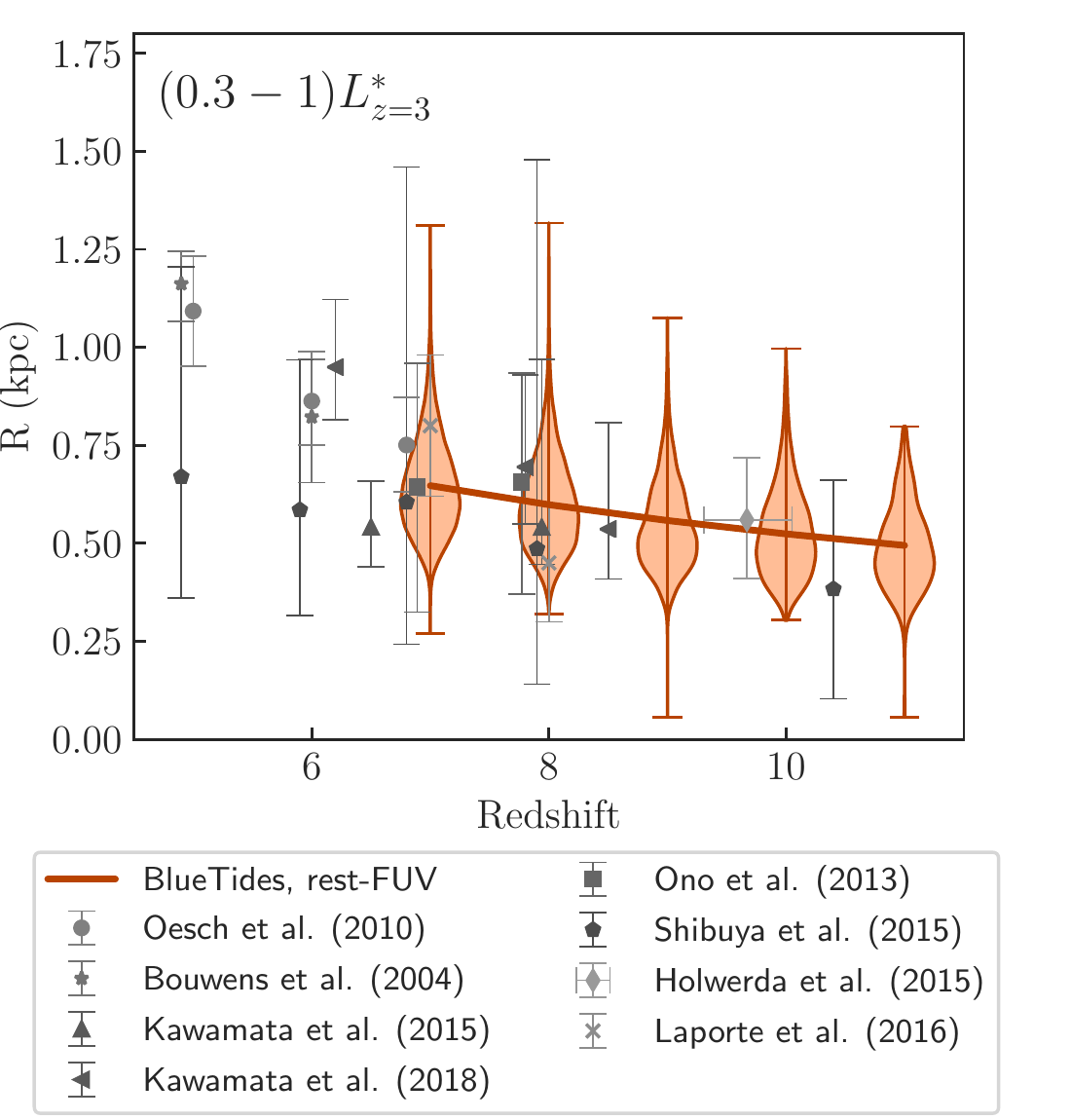}
\caption{The distribution of \BlueTides galaxy sizes at each redshift, alongside a best-fitting curve of the form $R\propto(1+z)^{-m}$. For this comparison, we consider galaxies with $(0.3-1)L^*_{z=3}$, or $L_{1500}=28.51$--29.03 erg/s/Hz, from the luminosity-limited sample. The limits on the violin plots show the 0.5\% and 99.5\% percentiles. Also shown are a range of high-redshift observations
\citep{Bouwens2004,Oesch2010,Ono2013,Shibuya2015,Holwerda2015,Laporte2016,Kawamata2015,Kawamata2018}.
}
\label{SizeRedshift}
\end{center}
\end{figure}

\subsubsection{Dependence on wavelength}
As the effect of dust is strongly wavelength dependent, the prediction that the size--luminosity relationship is driven predominantly by dust can be tested by observing the relationship at different rest-frame wavelengths. 
In Figure \ref{fig:wavelengths} we show the linear fit to the size--luminosity relationship for a range of bands stretching from the rest-frame FUV to near-infrared ($\simeq 2 \mu$m; see inset panel). 
As expected, the slope of the size--luminosity relation decreases at longer wavelengths where the impact of dust is smaller, becoming negative at the I-band and becoming more negative until the H-band. By the H-band, the size--luminosity relation almost mirrors the intrinsic FUV size--luminosity relation.

Observations at low redshift have found that galaxy sizes decrease at longer wavelengths \citep[e.g.][]{Barbera2010,Kelvin2012,Vulcani2014,Kennedy2015,Tacchella2015}, which is consistent with our prediction.
While it is not currently possible to probe the rest-frame optical or near-infrared of galaxies at high redshift, this will soon be possible with JWST, allowing us to test our theory that dust is the predominant driver of the positive size--luminosity relationship at high redshift.

\begin{table}
\begin{center}
\caption{The best-fitting dust-attenuated size--FUV luminosity relation in the form $R_e = R_0 \left({L_{1500}}/{L^\ast_{z=3}}\right)^\beta $ at each redshift from $z=7$ to $z=11$, for the luminosity-limited sample with $L_{1500}>10^{28.5}$ erg/s/Hz. }
\begin{tabular}{|c|c|c|c|c|}
\hline 
 &   
\multicolumn{2}{r}{Dust-Attenuated FUV} &    \multicolumn{2}{c}{Intrinsic FUV}\\
$z$ & $R_0$ (kpc) & $\beta$ & $R_0$ (kpc) & $\beta$ \\
\hline 
7 & $0.754\pm 0.001$ & $0.242 \pm 0.002$  & $0.515 \pm 0.001$ & $-0.087\pm0.001$  \\ 
8 & $0.671\pm 0.002$ & $0.166 \pm 0.003$  & $0.486 \pm 0.001$ & $-0.058 \pm 0.002$ \\ 
9 & $0.605\pm 0.003$ & $0.159 \pm 0.007$  & $0.439 \pm 0.001$ & $-0.073 \pm 0.005$ \\ 
10 & $0.567\pm 0.007$ & $0.12 \pm 0.02$ & $0.428 \pm 0.003$ & $-0.10 \pm 0.01$ \\ 
11 & $0.52\pm 0.02$ & $0.11 \pm 0.04$  & $0.394 \pm 0.007$ & $-0.12 \pm 0.03$  \\ 
\hline 
\end{tabular} 
\label{table}
\end{center}
\end{table}

\subsubsection{Redshift evolution}
We now consider the evolution of the FUV size--luminosity relation from $z=7$ to $z=11$. The best-fitting linear relation at each redshift for the luminosity-limited sample is given in Table \ref{table}, and shown in Figure \ref{LuminositySizeRedshift}. 
We find that the slope of the dust-attenuated relation decreases with redshift, with $\beta=0.242$ at $z=7$ and $\beta=0.11$ at $z=11$, with a decreasing normalization to higher redshifts. We find that the intrinsic relation remains negative from $z=11$ to $z=7$, with a slightly increasing normalization and roughly constant slope of $\beta\simeq-0.1$. These are both consistent with the mean size of galaxies increasing with redshift, as expected. The evolution in both relations is mild relative to the scatter in the distribution.

To investigate further, we consider the relation between galaxy size and redshift. In Figure \ref{SizeRedshift} we show the distribution of \BlueTides galaxy sizes at each redshift, alongside a best-fitting curve of the form $R\propto(1+z)^{-m}$, and a range of observations. For this comparison, we consider galaxies with $(0.3-1)L^*_{z=3}$, or $L_{1500}=28.51$--29.03 erg/s/Hz, from the luminosity-limited sample. We find that for $(0.3-1)L^*_{z=3}$ galaxies, $m=0.662\pm0.009$.
This is shallower than the values obtained by the observations shown in Figure \ref{SizeRedshift} which find $1\lesssim m \lesssim 1.5$ \citep{Bouwens2004,Oesch2010,Ono2013,Kawamata2015,Shibuya2015,Laporte2016,Kawamata2018}, except for \citet{Holwerda2015}, who predict $m=0.76\pm0.12$. Our value of $m$ is also smaller than the \citet{Liu2016} and \citet{Marshall2019} values from different versions of the \textsc{Meraxes} semi-analytic model, of $m=2.00\pm0.07$ and $m=1.98\pm0.07$ respectively, and the FIRE-2 hydrodynamical simulation results of \citet{Ma2018}, which predicted $m=1$--2, depending on the fixed mass or luminosity at which the relation is calculated. Both \textsc{Meraxes} and FIRE-2 included galaxy sizes from $z\geq5$ in their measurements of $m$, while we are constrained to $z\geq7$.
Their fits may be more heavily affected by the $z<7$ galaxy sizes, as this is a period of extreme galaxy growth where the sizes  increase rapidly, and so we may find closer agreement to these existing studies if \BlueTides was extended to lower redshifts. 
 
The \BlueTides sample exists only for $z\geq7$, beyond which only a few small samples of galaxy size measurements exist.  In addition, as mentioned above, our sizes are not calculated following an identical methodology to these observations. Thus, comparisons between observations and our predictions are difficult, and thus our shallow slope is not necessary an indication of a failure of our model. Sizes calculated using a more observational approach, and extending the simulation to lower redshifts, could yield significant differences in the extracted value of $m$.

\subsubsection{Comparison with other simulations}
\citet{Ma2018} studied the sizes of galaxies at $z=6$, 8 and 10 using FIRE cosmological
zoom-in simulations.
These high-resolution simulations have mass resolution for baryonic particles (gas
and stars) $M = 100-7000M_\odot$ and star particle softening lengths of 0.7–2.1 pc \citep[see also][]{Ma2018a}.
The majority of these galaxies have $M_\ast\lesssim10^{8}M_\odot$ and $M_{\textrm{UV}}\gtrsim-18$ mag, significantly less massive and fainter than those in our sample.  
To calculate the sizes of each galaxy, \citet{Ma2018} created stellar surface density and UV and B-band surface brightness images, excluding the effects of dust-attenuation. These images have a pixel size of 0.0032 arcsec, or 10--20 pc. 

\citet{Ma2018} found large scatter in the size--luminosity relation for these faint galaxies. As in our work, they found that redder wavelengths are better tracers of stellar mass than the UV emission.
They also found that galaxies appear smaller and more concentrated in the FUV compared to the B band and the intrinsic stellar mass distribution. This is in contrast to our results, which find that the galaxies appear larger in the FUV than in redder bands due to more dust attenuation at bluer wavelengths. \citet{Ma2018} do not implement dust attenuation on their images as the effect of dust would be negligible for these faint, low-mass galaxies, and so such an effect would not be seen.
Given the different modelling strategies, and particularly that the vast majority of these FIRE galaxies have luminosities well below those simulated by \BlueTidesns, making a direct comparison is difficult. \\
 
\citet{Wu2020} studied $\sim 5,000$ galaxies from the SIMBA cosmological hydrodynamic simulations at $z=6$, with $M_{1500}\gtrsim-21.5$ mag and $M_\ast\lesssim10^{10.5}M_\odot$. These are brighter than the \citet{Ma2018} simulated sample, overlapping with our \BlueTides luminosity distribution, although they are $\lesssim1.5$ mag fainter than our brightest galaxies. 
\citet{Wu2020} measured the sizes of galaxies from mock images at rest-frame 1600\AA~ and 6300\AA~ (between the V and I bands), from the JWST NIRCam F115W and F444W filters. To mimic JWST observations, the images have the NIRCam pixel scales of 0.031 and 0.063 arcsec respectively, which correspond to 0.18 and 0.37 kpc at $z = 6$. However, note that these images do not include other instrumental effects such as a PSF or noise. The images are created using a \citet{Calzetti2000} dust curve.
 
\citet{Wu2020} found a positive relation between the dust-attenuated FUV galaxy size and luminosity. They also found that the intrinsic galaxy sizes are a factor of 1.5--3 times smaller than the dust-attenuated sizes for galaxies with $M_\ast > 10^9 M_\odot$, with larger differences for more massive galaxies. This would result in a shallower intrinsic size--luminosity relation than the dust-attenuated relation.
This is broadly consistent with our finding that the effect of dust attenuation increases the sizes of high-redshift galaxies in the FUV, more significantly for the brightest and most massive galaxies, and is the cause of the positive size--luminosity relation in \BlueTidesns. 

The sizes of SIMBA galaxies are very similar in the rest-frame UV and optical, for both the intrinsic and dust-attenuated images \citep{Wu2020}. For similar luminosity $M_{1500}\gtrsim-21$ mag galaxies in our sample, we also find that the difference between FUV and optical sizes is small ($\lesssim0.15$ dex; Figure \ref{fig:wavelengths}). Thus by studying the brighter galaxies that form in \BlueTidesns, we are able to find that while the FUV and optical sizes are similar for faint, lower-mass galaxies, for the brightest galaxies dust attenuation results in larger FUV than optical galaxy sizes (Figure \ref{fig:wavelengths}).

\citet{Wu2020} found stronger dust attenuation in the centre of the galaxies.  They also found that the outskirts of some galaxies contain older, redder stars, while the centres are younger and bluer.  \citet{Wu2020} suppose that these age gradients counteract the effect of the dust on the colour gradient, resulting in similar rest-frame UV and optical sizes. We also see similar age gradients in lower mass/luminosity \BlueTides galaxies (Section \ref{sec:SizeMass} and Figure \ref{fig:FullImages}), with younger, brighter stars commonly found in the galaxy core. Indeed, for low luminosity $M_{\rm UV}>-20.5$ mag galaxies, we find equivalent galaxy sizes in all bands from the FUV to H-band (Figure \ref{fig:wavelengths}).
These age gradients may explain why \citet{Ma2018} see smaller FUV sizes than B-band sizes, in the absence of dust attenuation.
We note however that for the most massive and luminous \BlueTides galaxies, these age gradients are often reversed, with the younger, brighter stars instead found in the galaxy outskirts.

We note that the \citet{Wu2020} UV and optical images have different pixel sizes to mimic the JWST NIRCam instrument properties, which may affect the measured results. 
Both \citet{Ma2018} and \citet{Wu2020} considered only two bands, spanning a reduced wavelength range compared with that studied here.
Various image and size extraction methodologies in both \citet{Ma2018} and \citet{Wu2020} make a direct comparison difficult, before even considering the differences in the simulations themselves.

\section{Sizes of Quasar Host Galaxies}
\label{sec:Quasars}

\begin{figure*}
\begin{center}
\includegraphics[scale=0.7]{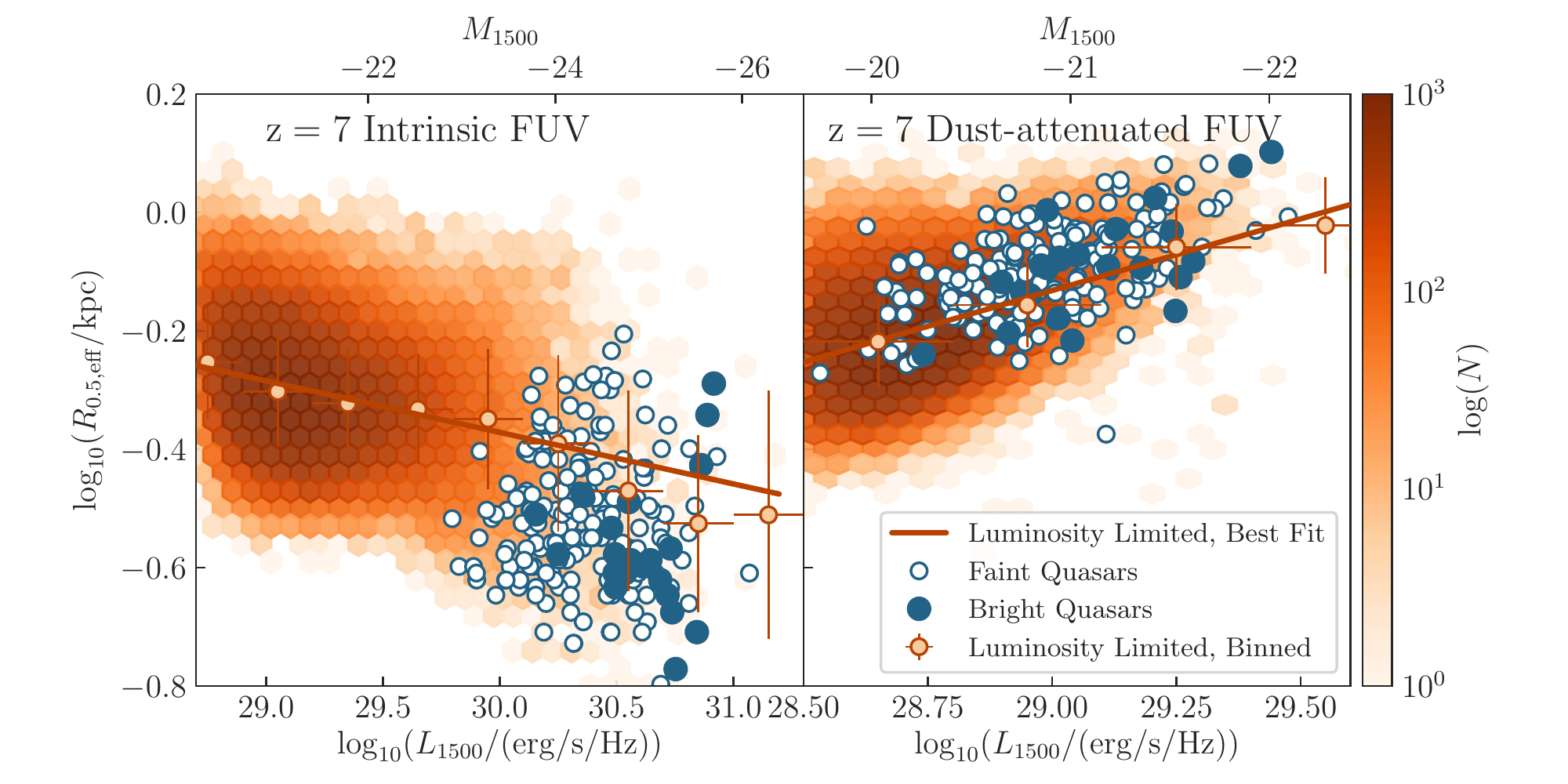}
\caption{The predicted relationship between intrinsic (left) and dust-attenuated (right) FUV luminosity ($\lambda=1500$\AA) and the measured size from that map. The sizes of both bright ($m_{\textrm{UV,AGN}}<22.8$ mag) and faint ($22.8<m_{\textrm{UV,AGN}}<24.85$ mag) quasars are shown (see legend).
The orange density plot depicts all galaxies with dust-attenuated $L_{1500}>10^{28.5}$ erg/s/Hz at $z = 7$, i.e. the luminosity-limited sample. The orange line shows the linear best fit to this distribution.  The orange points show the median for the luminosity-limited sample in bins of 0.3 dex, for bins with more than 10 galaxies, with vertical errorbars depicting the standard deviation of the distribution.}
\label{fig:QuasarSizeLuminosity}
\end{center}
\end{figure*}

The large volume of \BlueTides contains a statistical sample of high-redshift quasars, not produced at these redshifts by smaller simulations such as IllustrisTNG \citep[e.g.][]{Pillepich2017,Weinberger2018}, Eagle \citep[e.g.][]{Schaye2014,McAlpine2017}, 
Horizon-AGN \citep[e.g.][]{Volonteri2016,Habouzit2019b}, and SIMBA \citep[e.g.][]{Dave2019}. The properties of \BlueTides $z=7$ quasars were analysed in detail in \citet{Marshall2020}. In that work, we found that the one distinguishing feature of quasar host galaxies relative to the overall $z=7$ galaxy sample was their half-mass radii, with quasar hosts having compact half-mass radii of $0.40\substack{+0.11 \\ -0.09}$ kpc, whereas galaxies of similar mass and luminosity have a wider range of sizes with a larger median value, $R_{0.5}=0.71\substack{+0.28 \\ -0.25}$ kpc.

This analysis was performed using the stellar half-mass radius, calculated from the 3D star particle distribution. In \citet{MarshallBTpsfMC} we created mock JWST images of these quasar hosts, and found that these half-mass radii were not well correlated with the S\'ersic radii measured from the mock galaxy images. This is likely due to the  half-mass radius being calculated from the stellar particle distribution, which does not consider the dust distribution nor any variation in the
mass-to-light ratio. Thus, while \citet{Marshall2020} predict that the half-mass radii of quasar host galaxies are compact, this may not be reflected in their observed sizes.

With our detailed size measurements, here we have the ability to perform a more robust analysis of the sizes of quasar host galaxies relative to the overall galaxy population. 
As in \citet{Marshall2020} and \citet{MarshallBTpsfMC}, we define two quasar samples:
\begin{itemize}[leftmargin=*,noitemsep,nolistsep]
\item Bright quasars: 
$M_{\textrm{UV,AGN}}<M_{\textrm{UV,Host}}$ and $m_{\textrm{UV,AGN}}<22.8$ mag
\item Faint quasars: 
$M_{\textrm{UV,AGN}}<M_{\textrm{UV,Host}}$ and $22.8<m_{\textrm{UV,AGN}}<24.85$ mag
\end{itemize}
This assumes that `quasars' are AGN which outshine their host galaxy in the UV-band, i.e. $M_{\textrm{UV,AGN}}<M_{\textrm{UV,Host}}$, where $M_{\textrm{UV,AGN}}$ and $M_{\textrm{UV,Host}}$ are the observed (dust-attenuated) UV absolute magnitudes for the AGN and host galaxy, respectively.
The limiting magnitudes are taken from the faintest known SDSS quasar \citep[SDSS J0129--0035 with $m_{1450}=22.8$ mag;][]{Wang2013,Banados2016} 
and Subaru High-z Exploration of Low-Luminosity Quasars (SHELLQs) survey quasar \citep[HSC J1423--0018 with $m_{1450}=24.85$ mag;][]{Matsuoka2018}.

\begin{figure*}
\begin{center}
\includegraphics[scale=0.7]{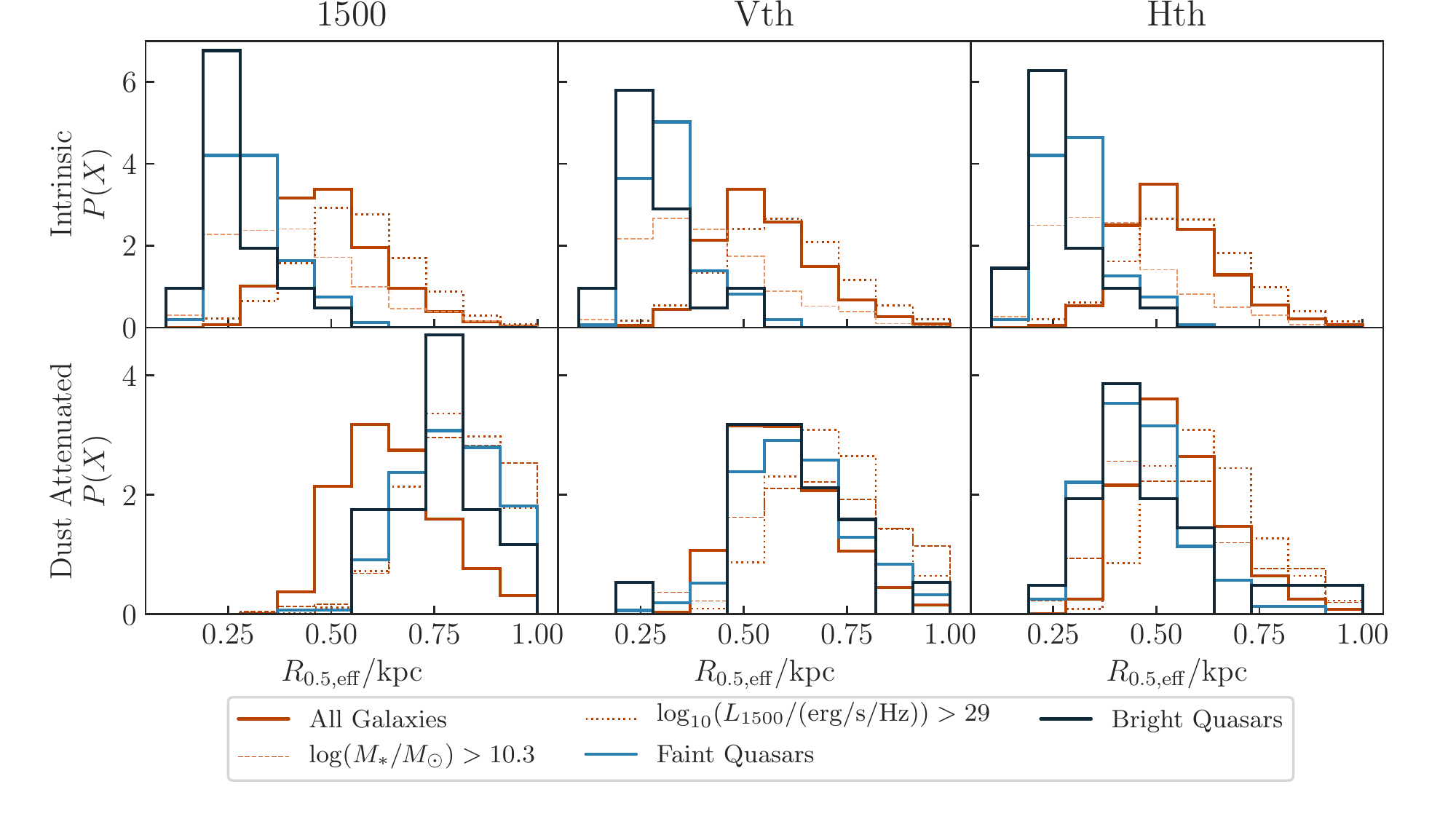}
\caption{The probability density of effective radius, for bright quasars, faint quasars, all galaxies, galaxies with similar mass to the quasars ($\log{M_\ast/M_\odot}>10.3$), and galaxies with similar dust-attenuated FUV luminosities to the quasars, ($L_{1500}>10^{29}$ erg/s/Hz; see legend). The sizes are calculated from the intrinsic (top rows) and dust-attenuated (bottom rows) maps, in the FUV 1500\AA~ (left), V (middle) and H (right) bands.}
\label{fig:QuasarSizeHistograms}
\end{center}
\end{figure*}

The \BlueTides simulation contains 22 bright quasars and 175 faint quasars at $z=7$, with the brightest quasar in the simulation having $m_{\textrm{UV,AGN}}=20.7$ mag. For full details, see \citet{Marshall2020}.
In Figure \ref{fig:QuasarImages} we show V-band images of six quasar host galaxies, as well as six non-quasar hosts in the simulation with similar stellar masses and luminosities, both with and without dust attenuation. 
Note that we do not include the AGN emission in our images, to allow for accurate measurements of the host galaxy.

In Figure \ref{fig:QuasarSizeLuminosity}, we show the intrinsic and dust-attenuated size--luminosity relations at $z=7$ for the bright and faint quasars, alongside the luminosity-limited galaxy sample. As in \citet{Marshall2020}, we find that the intrinsic sizes of quasar host galaxies are more compact than galaxies with similar dust-attenuated FUV luminosities: bright quasars have effective radii $0.26 \substack{{+0.07}\\{-0.04}}$ kpc, faint quasars have radii $0.30 \substack{{+0.10}\\{-0.06}}$ kpc, and galaxies with $L_{1500}>10^{29}$ erg/s/Hz have radii $0.56 \substack{{+0.14}\\{-0.13}}$ kpc, measured from the intrinsic FUV maps. 
However, we find that the quasar host sizes are \textit{not} small, and are often even larger than the average galaxy, when the size is measured from the dust-attenuated FUV maps: bright quasars have radii $0.81 \substack{{+0.17}\\{-0.13}}$ kpc, faint quasars have radii $0.80 \substack{{+0.16}\\{-0.13}}$, and galaxies with $L_{1500}>10^{29}$ erg/s/Hz have radii $0.82 \substack{{+0.14}\\{-0.12}}$, measured from the dust-attenuated FUV maps. 

In Figure \ref{fig:QuasarSizeHistograms} we show the probability density distributions of bright quasars, faint quasars, all galaxies, galaxies with similar mass to the quasars, $\log{M_\ast/M_\odot}>10.3$, and galaxies with similar dust-attenuated FUV luminosities to the quasars, $L_{1500}>10^{29}$ erg/s/Hz. We show these for the radii calculated from both the intrinsic and dust-attenuated maps, in the FUV, V and H bands.
We see that in all three bands, the intrinsic sizes of quasars are small compared to the overall galaxy samples, and bright quasars show more compact sizes than faint quasars. However, the dust-attenuated sizes are similar to the overall samples in all three filters.

The half-mass radii \citep{Marshall2020} and effective radii measured from intrinsic (dust-free) images of quasars are small relative to the overall galaxy population at $z=7$. This implies some physical mechanism which results in quasar hosts being physically smaller than their non-quasar counterparts. \citet{Matteo2017} found that in \BlueTidesns, effective black hole growth occurs in regions with low tidal fields, where cold gas is accreted along thin, radial filaments straight into the centre of the halo, forming the most compact galaxies and most massive black holes at the earliest times. Using constrained Gaussian realizations, \citet{Ni2020} also found that a highly compact initial density peak and low tidal field are favourable for forming high density gas environments, which result in massive black holes and compact galaxy morphologies.
This is also consistent with observations at low redshift, where sizes of quasar hosts are generally more compact than star-forming galaxies of equivalent stellar mass, but not as compact as quiescent galaxies \citep{Silverman2019,Li2021}; this is hypothesised to be due to AGN being preferentially triggered within more compact galaxies.

However, when measuring galaxy sizes in UV images which include dust-attenuation, the \BlueTides quasar hosts are no longer small, due to the large amount of dust attenuation resulting in a flatter emission distribution  (see e.g. Figure \ref{fig:QuasarImages}). Thus, these physical trends of quasar hosts being compact are unlikely to be detectable with rest frame UV imaging. The upcoming era of quasar host observations with JWST, for example, may therefore be unable to test our theoretical expectations that quasar host galaxies are more compact, and thus these results must be carefully considered when such observations are interpreted.

\section{Conclusions}
\label{sec:Conclusions}
In this work, we used the large cosmological hydrodynamical simulation \BlueTides to investigate the sizes of $\sim$100,000 simulated galaxies in the Epoch of Reionisation ($7 \leq z \leq 11$).
We considered a mass-limited sample of galaxies with $M_\ast>10^{8.5}M_\odot$, and a luminosity-limited sample of galaxies with dust-attenuated FUV (1500\AA) luminosity $L_{1500}>10^{28.5} {\textrm{erg/s/Hz}}$. 

We created rest-frame UV and optical images of each galaxy in standard top-hat filters, both with and without dust attenuation to explore the effect of dust on the measured galaxy sizes.
We then measured the effective galaxy sizes from these images using a similar methodology to \citet{Ma2018}.

Our conclusions are:
\begin{itemize}[leftmargin=*,nolistsep]
\item There is an inverse relation between size (as measured from the stellar mass map) and stellar mass, suggesting that the most massive galaxies are more compact and dense than lower mass galaxies, which have flatter mass distributions (see Figures \ref{fig:FullImages}, \ref{fig:stellarMassSize} and \ref{fig:surfaceDensities}). 
\item The relation between intrinsic FUV size and luminosity is slightly negative, with a median galaxy size of $\sim 0.5$ kpc (Figure \ref{fig:FUVSize}). 
\item The dust-attenuated FUV size--mass relation shows very different properties to the intrinsic relation. Sizes \textit{increase} with FUV luminosity, with $R_e\propto L^{0.242}$ for the luminosity-limited sample at $z=7$ (Figure \ref{fig:FUVSize}). Thus, the positive FUV size--mass relation is
driven predominantly by the effects of dust. Dust preferentially attenuates brighter sight lines, resulting in flatter emission profiles (see Figures \ref{fig:FullImages} and \ref{fig:surfaceDensities}).
\item Our predicted dust-attenuated FUV size--luminosity relation has a slope that is shallower than current observational estimates. However, we must be cautious in such comparisons as we apply a different object selection and size measurement methodology to observations.
\item Because dust attenuation is a strong function of wavelength, the slope of the size--luminosity relation decreases and becomes negative at rest-frame optical/near-infrared wavelengths (Figure \ref{fig:wavelengths}). This raises the possibility of testing our conclusion with JWST, which will provide high-resolution rest-frame UV and optical imaging of galaxies at high redshift.
\item The slopes of the FUV size--luminosity relations evolve slightly from $z=11$ to 7, while their normalization increases (Figure \ref{LuminositySizeRedshift}). This evolution is mild relative to the expected scatter in the distribution.
\item For $(0.3-1)L^*_{z=3}$ galaxies, their size evolves with redshift as $R\propto(1+z)^{-m}$, where $m=0.662\pm0.009$. This is shallower than current observational measurements and theoretical predictions, albeit a direct comparison is difficult.
\item Quasar host galaxies are small relative to similar galaxies when their sizes are measured from intrinsic FUV images. However, we find that the quasar host sizes are \textit{not} small, when the size is measured from dust-attenuated FUV images, due to the large amount of dust in their hosts. Thus, while quasar hosts are predicted to be compact due to galaxy formation physics, this prediction is unlikely to be testable with rest-frame UV observations of quasar hosts.\\
\end{itemize}

In this work we focus on a simple investigation of galaxy sizes, making mock images with standard top-hat filters in the rest-frame UV and optical. For a more comprehensive comparison with existing and upcoming observations, ideally one would include observational effects such as noise, surface brightness and magnitude limits, and the instrument pixel size, resolution, filters and PSF. A source finder would then be run on the images, with properties extracted using the same method as for real observations. While beyond the scope of this paper, we plan in future work to make detailed predictions for the galaxy sizes measured with JWST, for example, to see how such instrumental effects would alter our conclusions.

\section*{Acknowledgements}
We thank the anonymous referee for useful feedback which improved this manuscript.
The \BlueTides simulation was run on the BlueWaters facility at the National Center for Supercomputing Applications.
Part of this work was performed on the OzSTAR national facility at Swinburne University of Technology, which is funded by Swinburne University of Technology and the National Collaborative Research Infrastructure Strategy (NCRIS).
MAM acknowledges the support of a National Research Council of Canada Plaskett Fellowship, and the Australian Research Council Centre of Excellence for All Sky Astrophysics in 3 Dimensions (ASTRO 3D), through project number CE170100013.
TDM acknowledges funding from NSF ACI-1614853, NSF AST-1517593, NSF AST-1616168 and NASA ATP 19-ATP19-0084 and 80NSSC20K0519,ATP.
TDM and RAC also acknowledge ATP 80NSSC18K101 and NASA ATP 17-0123.

This paper made use of Python packages and software 
AstroPy \citep{Astropy2013},
BigFile \citep{Feng2017},
FLARE \citep{FLARE},
Matplotlib \citep{Matplotlib2007},
NumPy \citep{Numpy2011},
Pandas \citep{reback2020pandas}, 
Photutils \citep{photutils},
SciPy \citep{2020SciPy-NMeth},  
and SynthObs \citep{SynthObs}.
This paper also makes use of version 17.00 of Cloudy, last described by \citet{Ferland2017}, and version 2.2.1 of the Binary Population and Spectral Population Synthesis (BPASS) model \citep{Stanway2018}.

\section*{Data Availability}
Data of the \textsc{BlueTides} simulation is available at \url{http://bluetides.psc.edu}.
The data generated in this work will be shared on reasonable request to the corresponding author.

\bibliographystyle{mnras}
\bibliography{BTSizesPaper.bib} 


\appendix

\section{The effect of model choices on galaxy size}
\subsection{Dust models}
\label{App:Dust}

\begin{figure*}
\begin{center}
\includegraphics[scale=0.7]{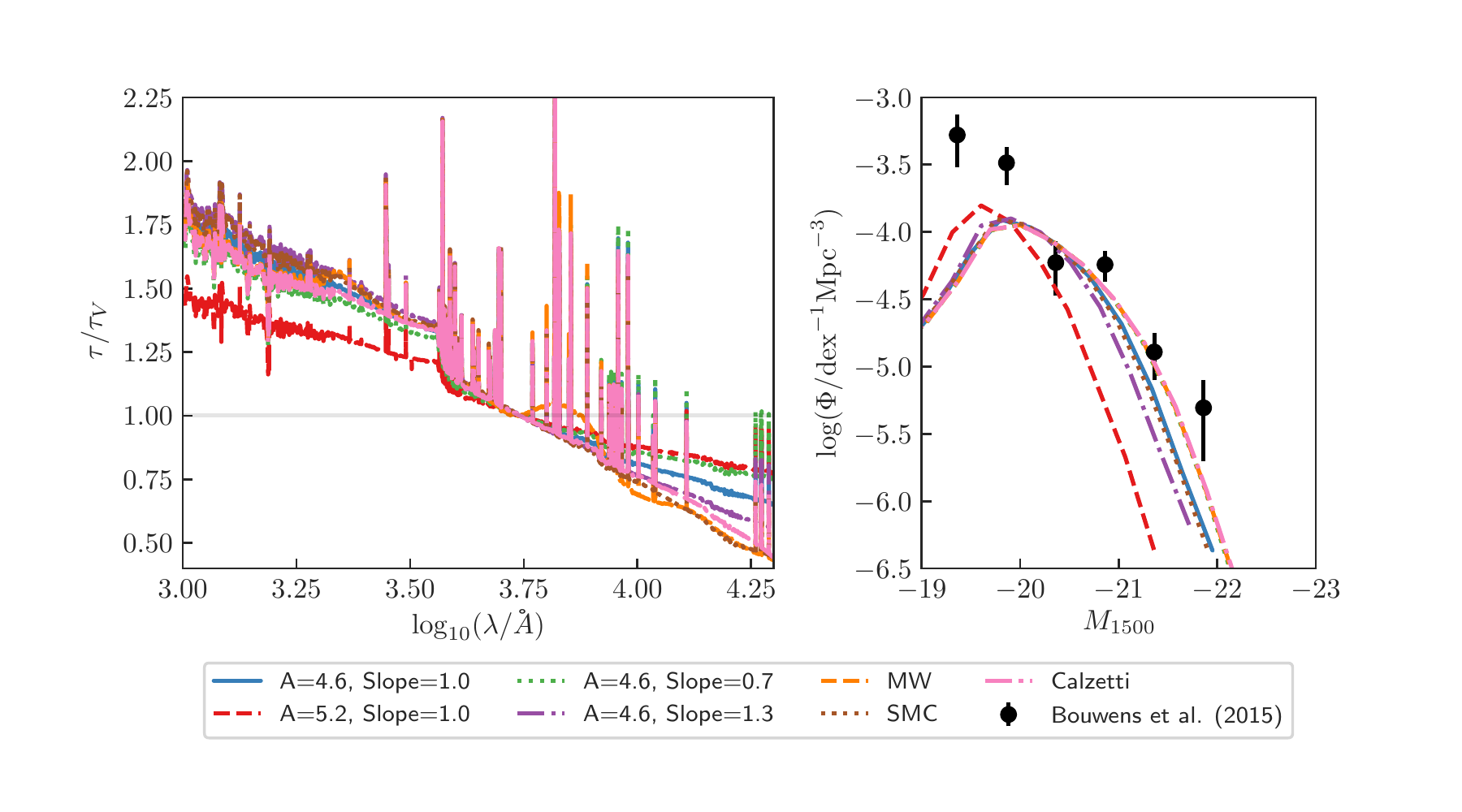}
\caption{
The attenuation curves for an example galaxy (left), and the resulting $z=8$ UV galaxy luminosity functions (right) for the various dust models considered in Appendix \ref{App:Dust}, compared with the observations of \citet{Bouwens2015}.  
}
\label{fig:Dust}
\end{center}
\end{figure*}

Throughout this work we assume a dust attenuation law with
$\tau_\lambda \propto \lambda^{-1}$, i.e. $\gamma=-1$, and $\kappa=10^{4.6}$, which is calibrated against the observed galaxy UV luminosity function at redshift $z=7$. 

Here we consider the effect of using different dust attenuation laws on the measured galaxy sizes.
We consider four basic relations: the standard $\gamma=-1$ and $\kappa=10^{4.6}$, a more severe dust attenuation with $\gamma=-1$ and $\kappa=10^{5.2}$, and different slopes $\gamma=-1.3$ and $\kappa=10^{4.6}$, and $\gamma=-0.7$ and $\kappa=10^{4.6}$. We also consider the Milky Way (MW) and Small Magellanic Cloud (SMC) attenuation curves from \citet{Pei1992}, as well as the attenuation curve from \citet{Calzetti2000}, using $\kappa=10^{4.6}$.
The various attenuation curves and the resulting $z=8$ galaxy luminosity functions are shown in Figure \ref{fig:Dust}. 

\begin{figure*}
\begin{center}
\includegraphics[scale=0.7]{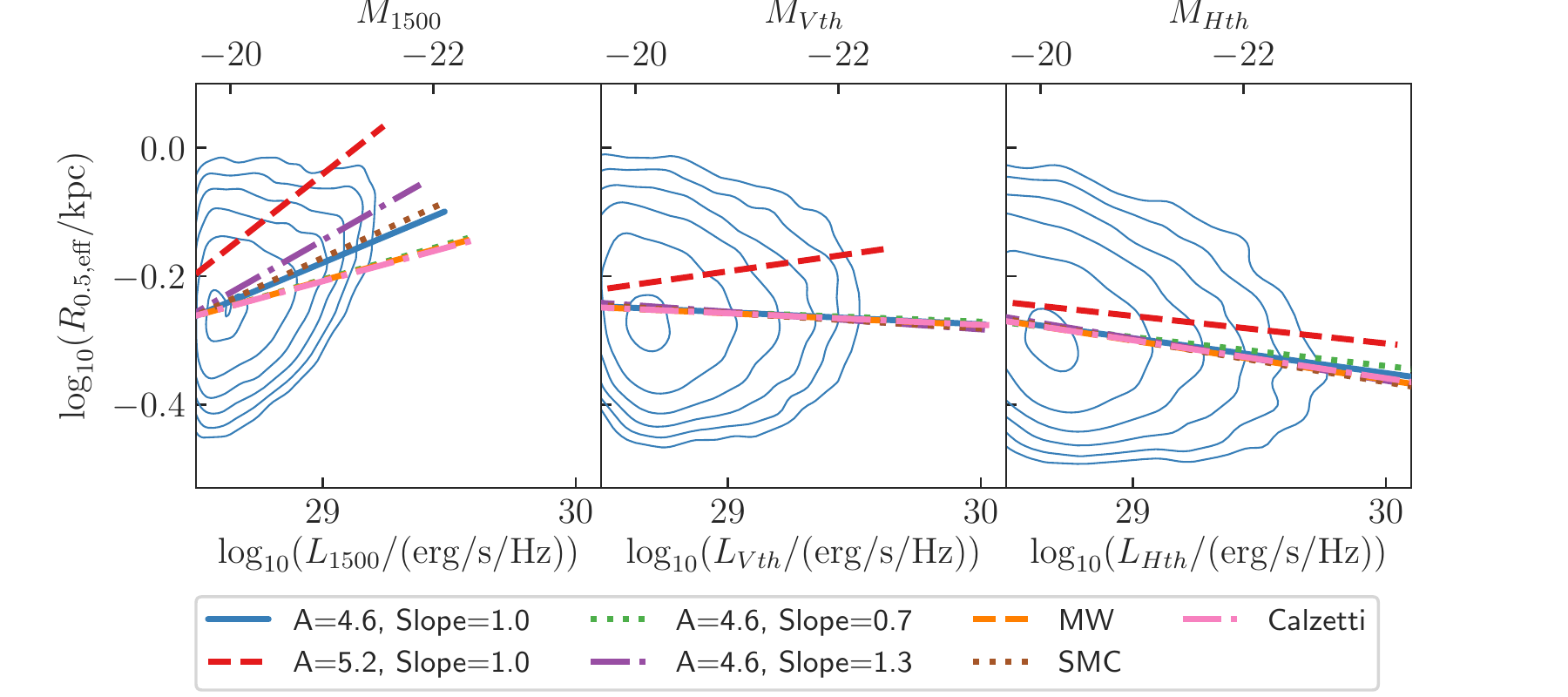}
\includegraphics[scale=0.7]{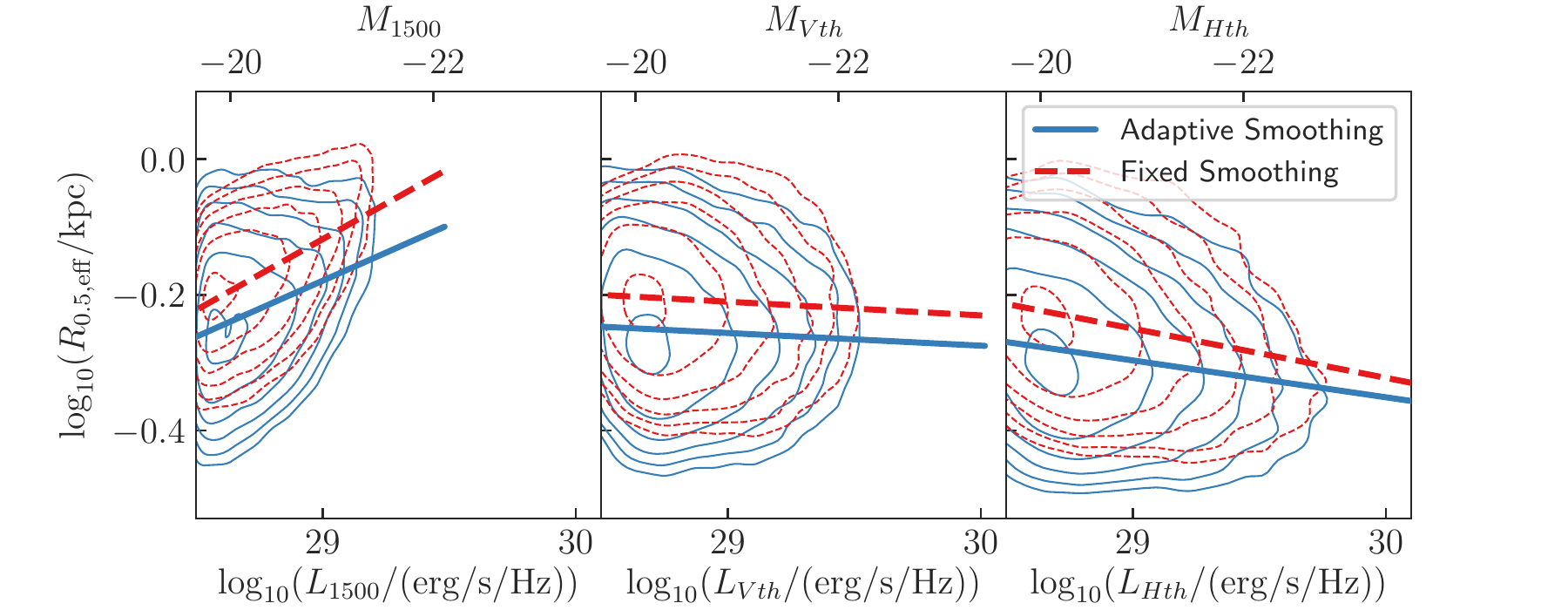}
\includegraphics[scale=0.7]{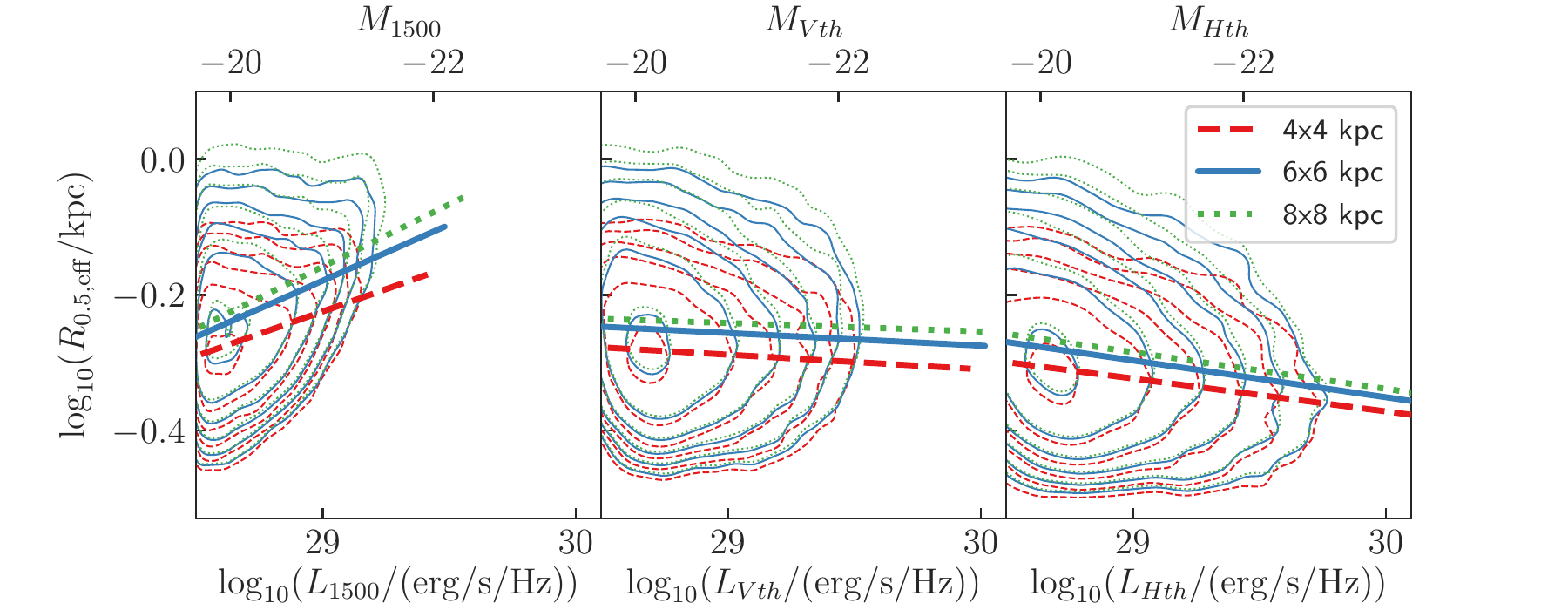}
\includegraphics[scale=0.7]{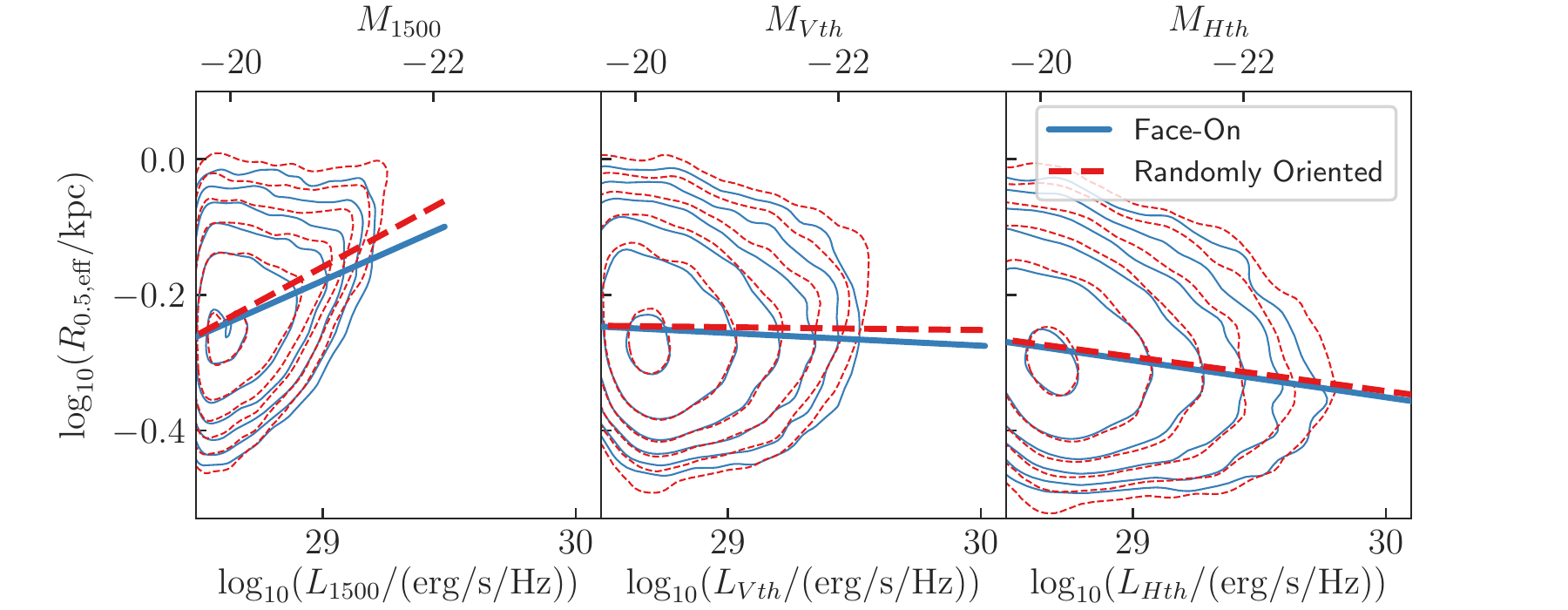}
\caption{The different $z=8$ size--luminosity relations measured in the FUV 1500\AA~ (left column), V (middle column) and H (right column) filters under various model assumptions. 
Top row: the various dust models considered in Appendix \ref{App:Dust},
Second row: the adaptive and fixed smoothing length assumptions,
Third row: assuming an image FOV of $4\times4$, $6\times6$ and  $8\times8$ kpc,
and Bottom row: assuming a face-on or random orientation.
Lines depict the best-fitting linear relation, while the contours depict regions containing 10\%, 55\%, 77\%, 88\%, 94\%, and 97\% of galaxies for the corresponding samples.}
\label{fig:AppendixRelations}
\end{center}
\end{figure*}

The $z=8$ size--luminosity relations for each model are shown in Figure \ref{fig:AppendixRelations}.
For the various $\kappa=10^{4.6}$ curves, the relations vary by a maximum of $\sim0.1$ dex in the FUV 1500\AA~ band, and the variation is minimal ($\lesssim0.02$ dex) in the V and H bands.
With more dust attenuation, the $\kappa=10^{5.2}$ model results in a smaller number of galaxies in the luminosity-limited sample, around $\sim70\%$ of the sample size of the standard $\gamma=-1$ and $\kappa=10^{4.6}$ model. This $\kappa=10^{5.2}$ model shows the largest variation in the size--luminosity relations from the other dust models, with the difference $\lesssim 0.25$ dex in the FUV. The variation is more pronounced at the largest luminosities, with the relation showing an increased slope.
This dust model shows a lower luminosity function, which significantly underestimates the \citet{Bouwens2015} observations (Figure \ref{fig:Dust}). Thus the $\kappa=10^{5.2}$ model produces excessive dust attenuation, with the $\kappa=10^{4.6}$ models providing a better match to the observed luminosity function.
Overall, these various $\kappa=10^{4.6}$ dust models all result in a small $\lesssim 0.1$ dex difference in the size--luminosity relations, and do not affect our main conclusions.

\subsection{Image smoothing}
\label{App:Smoothing}
When creating the galaxy images, we smooth the light of each star particle on to a 0.1 kpc grid. Throughout this work, this smoothing is performed adaptively, with the smoothing scale for each particle (full width at half maximum of the Gaussian) set to the distance to the 8th nearest neighbour (in 3D). 

\begin{figure*}
\begin{center}
\includegraphics[scale=0.8]{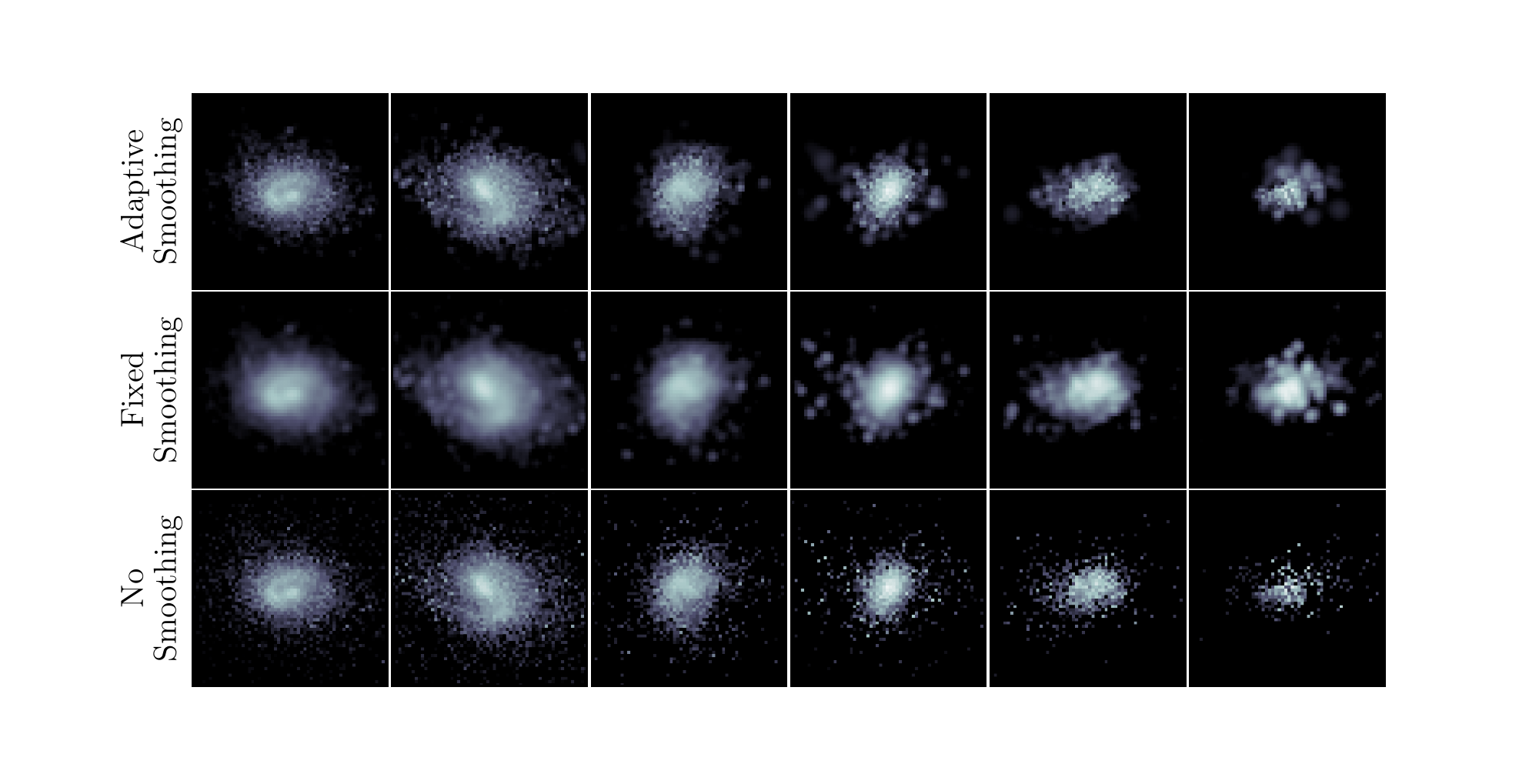}
\caption{Images of six $z=8$ galaxies in the V-band under different smoothing assumptions: adaptively smoothing based on the distance to the 8th nearest neighbour as used throughout this work (top row),  a fixed smoothing scale equal to the gravitational softening length $1.5/h/(1+z)$ pkpc (middle row), and no smoothing (bottom row). The images show a $6\times6$ kpc FOV. The galaxies are those shown in 
Figure \ref{fig:FullImages}. In the standard, adaptive smoothing images the galaxies have V-band effective radii of (left-to-right) 5.86, 7.18, 5.41, 4.62, 4.44, and 4.03 kpc, 
with fixed smoothing they have radii of 6.13, 7.46, 5.70,  4.98, 5.01, and 4.58 kpc, and with no smoothing they have radii of 5.86, 7.16, 5.38, 4.55, 4.33, and 2.93 kpc.
}
\label{fig:AppendixSmoothingImages}
\end{center}
\end{figure*}

Here, we consider instead assuming a fixed smoothing scale, equal to the gravitational softening length $1.5/h/(1+z)$ pkpc.
Images of $z=8$ galaxies in the V-band under these two smoothing assumptions, as well as images with no smoothing, are shown in Figure \ref{fig:AppendixSmoothingImages}.
The fixed smoothing length provides a much more severe smoothing than the adaptive smoothing prescription.
We find that for the $z=8$ galaxy sample, the median $R_{\mathrm{fixed}}/R_{\mathrm{adaptive}}$ is $1.22\substack{{+0.08}\\{-0.10}}$ at 1500\AA, $1.14\substack{{+0.03}\\{-0.05}}$ in V and $1.13\substack{{+0.03}\\{-0.04}}$ in H, where errors correspond to the 16th and 84th percentiles. The fixed smoothing therefore results in larger measured sizes, whereas the adaptive smoothing allows finer details to be resolved and thus smaller sizes are generally measured.
In Figure \ref{fig:AppendixRelations} we show the different size--luminosity relations measured using these two smoothing assumptions, in the 1500\AA, V and H filters. We find that while the fixed smoothing produces larger sizes, this makes only a small difference ($\lesssim 0.1$ dex) to the overall normalisation of the luminosity--size relation and does not affect our main conclusions.

\subsection{Image field of view}
\label{App:FOV}
Throughout this work we consider images with a $6\times6$ kpc FOV.
Smaller images can exclude more extended emission from the galaxies, resulting in lower measured radii.
We find that the measured sizes generally converge on a value at a FOV of $\sim6\times6$ kpc. 
In Figure \ref{fig:AppendixRelations} we show the size--luminosity relation as measured from $4\times4$ kpc, $6\times6$ kpc and $8\times8$ kpc images. The $4\times4$ kpc images measure smaller sizes, with the relation up to $\sim0.06$ dex smaller for the brightest galaxies in the FUV, with a less significant effect in the V and H bands. A larger difference in the FUV is expected due to the larger dust attenuation resulting in flatter images, with the extended emission that is excluded from these smaller images contributing a higher portion of the total flux. Conversely, the $8\times8$ kpc images show very similar size--luminosity relations to the $6\times6$ kpc images in all bands, with difference $\lesssim 0.03$ dex. This indicates that the $6\times6$ kpc images do not exclude significant fractions of the galaxies and thus we are justified in considering these smaller, less computationally expensive images. 
Thus, we decide to use $6\times6$ kpc images throughout this work. This is consistent with the 1 arcsec diameter aperture used by \citet{Ma2018}.
This choice does not significantly change our results, with only a $\lesssim 0.03$ dex decrease in the size--luminosity relation relative to $8\times8$ kpc images.

\subsection{Galaxy orientation}
\label{App:Orientation}
The images throughout this work are created assuming the galaxy is seen from the `face-on' direction, where the normal vector of the image plane is aligned to the total angular momentum of the galaxy.
Here we consider the effect of this assumption on the measured galaxy sizes, by creating images with a random orientation. 
For this, we assume the galaxies are all viewed in the $z$-direction in the \BlueTides cube.
Note that the `face-on' direction is determined by the motion of the particles and their moment of inertia, which does not necessarily correspond to the direction at which the galaxy is largest.

\begin{figure*}
\begin{center}
\includegraphics[scale=0.8]{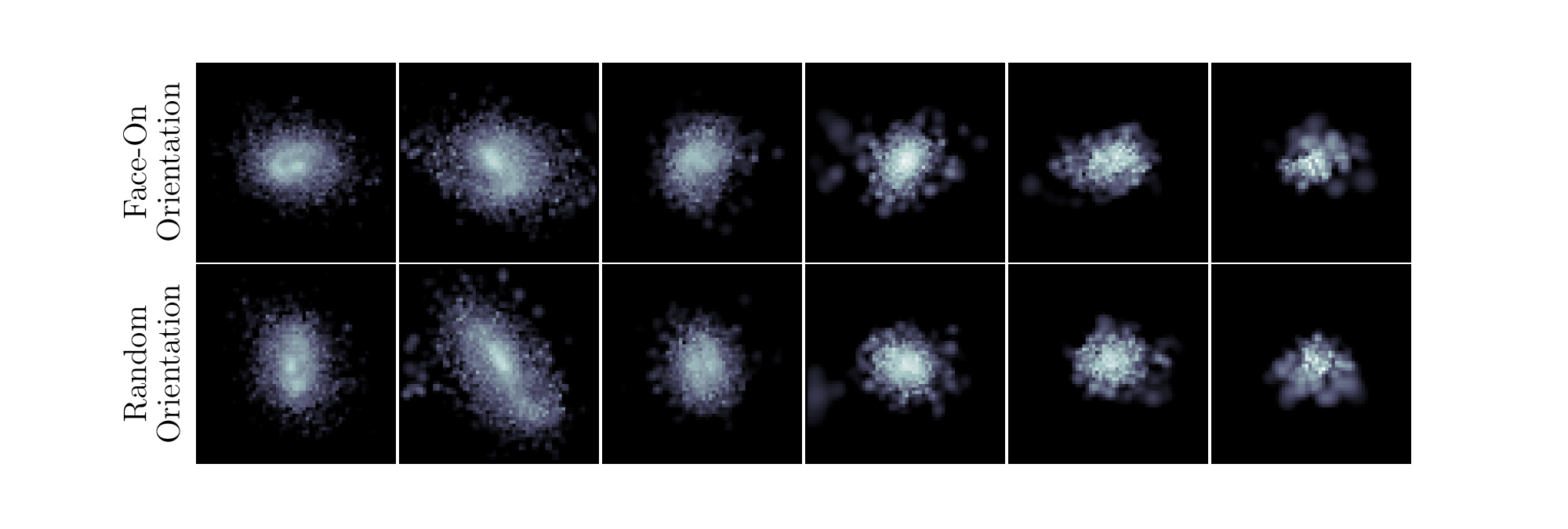}
\caption{Images of six $z=8$ galaxies in the V-band under the different orientation assumptions: face-on orientation, as used throughout this work (top row), and random orientation (bottom row). The images show a $6\times6$ kpc FOV. The galaxies are those shown in 
Figure \ref{fig:FullImages}. In the face-on direction the galaxies have V-band effective radii of (left-to-right) 5.86, 7.18, 5.41, 4.62, 4.44, and 4.03 kpc, 
and in the random orientation they have radii of 6.02, 7.84, 5.44, 4.89, 4.69, and 4.41 kpc.
}
\label{fig:AppendixOrientationImages}
\end{center}
\end{figure*}

Images of $z=8$ galaxies in the face-on and random orientations are shown in Figure \ref{fig:AppendixOrientationImages}, for comparison. 
In Figure \ref{fig:AppendixRelations} we show the size--luminosity relations measured for these different orientations, in the 1500\AA, V and H filters. 
Overall, the 
orientation makes only a small difference ($< 0.04$ dex) to the luminosity--size relation and does not affect our main conclusions.

\section{A comparison of size measurement techniques}
\label{App:SizeMeasurements}

Throughout this work we define the radius of each galaxy as its effective size:  $r_e=\sqrt{A_e/\pi}$, where $A_e$ is the minimum area encompassing 50\% of the total light of galaxy, even if the contributing pixels are non-contiguous. 
This approach is well-suited to clumpy, irregular high-redshift galaxies \citep{Ma2018}. 

\begin{figure*}
\begin{center}
\includegraphics[scale=0.7]{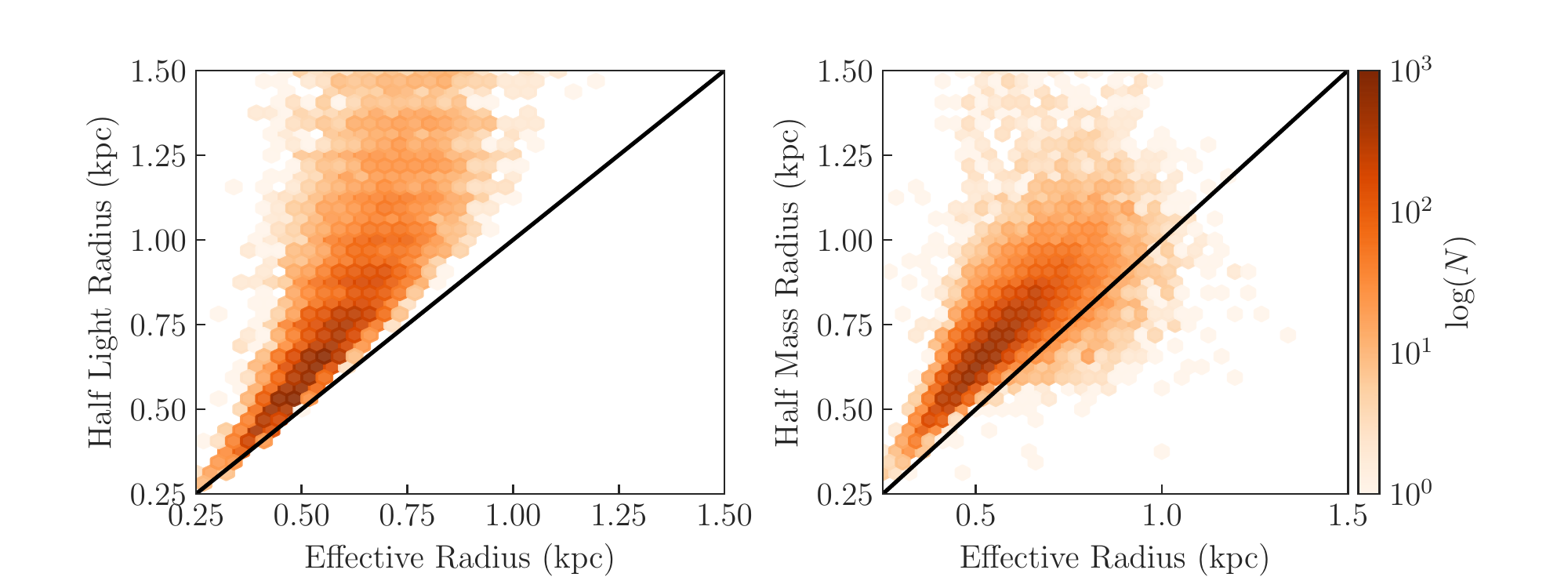}
\caption{Left: The galaxy half-light radius, measured using circular aperture photometry on the stellar mass maps, and Right: the half-mass radius inside
$R_{200}$ based on the 3D distribution of star particles from \BlueTidesns, compared to the
effective radius measured from the stellar mass maps at $z=8$. Lines show the 1:1 relation.}
\label{fig:AppRadiusComparison}
\end{center}
\end{figure*}

However, a more common technique for measuring galaxy sizes is curve-of-growth analysis, which constructs light profiles using circular or elliptical apertures 
to measure the half-light radius.
These are useful for measuring the sizes of smoothly varying light distributions with well defined centres. 
We follow this technique by performing aperture photometry in 20 circular apertures with radii log-spaced from 0.05 to 2 kpc, and interpolating the counts to find the half-light radius.
We centre the photometry on the brightest pixel in the image.
We note that choosing the intensity-weighted centre instead may produce different results.
In Figure \ref{fig:AppRadiusComparison} we compare this half-`light' radius to the effective radius, as measured on the stellar mass maps. 
We find that majority of the half-light radii are in reasonable agreement with the effective radii, although the half-light radii are generally larger. 
Equivalent trends are seen from the luminosity maps.

\begin{figure*}
\begin{center}
\includegraphics[scale=0.7]{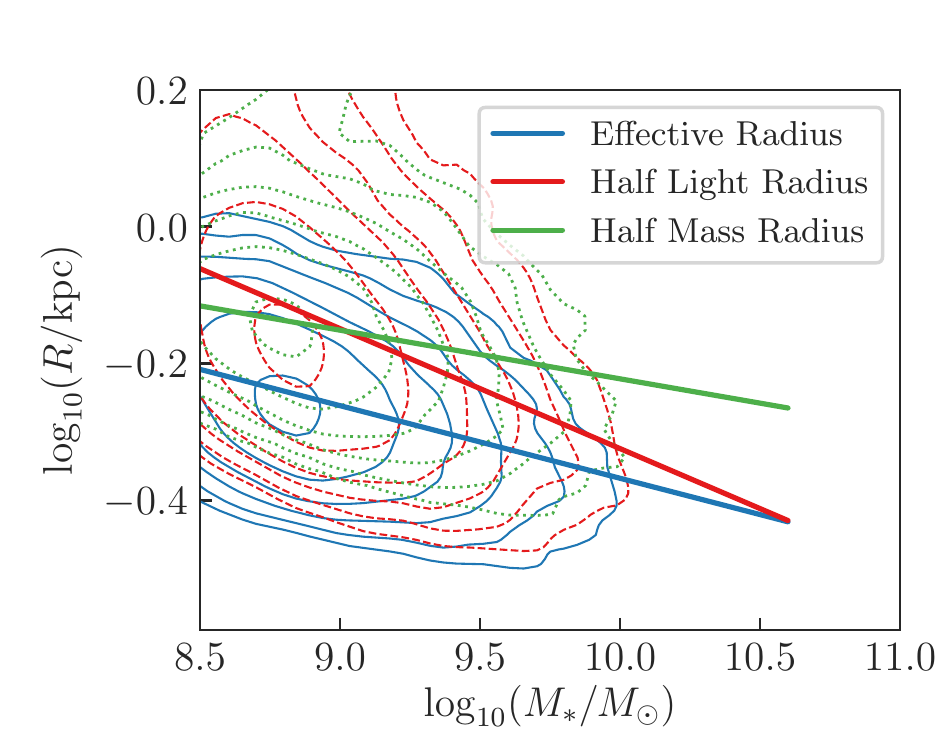}
\includegraphics[scale=0.7]{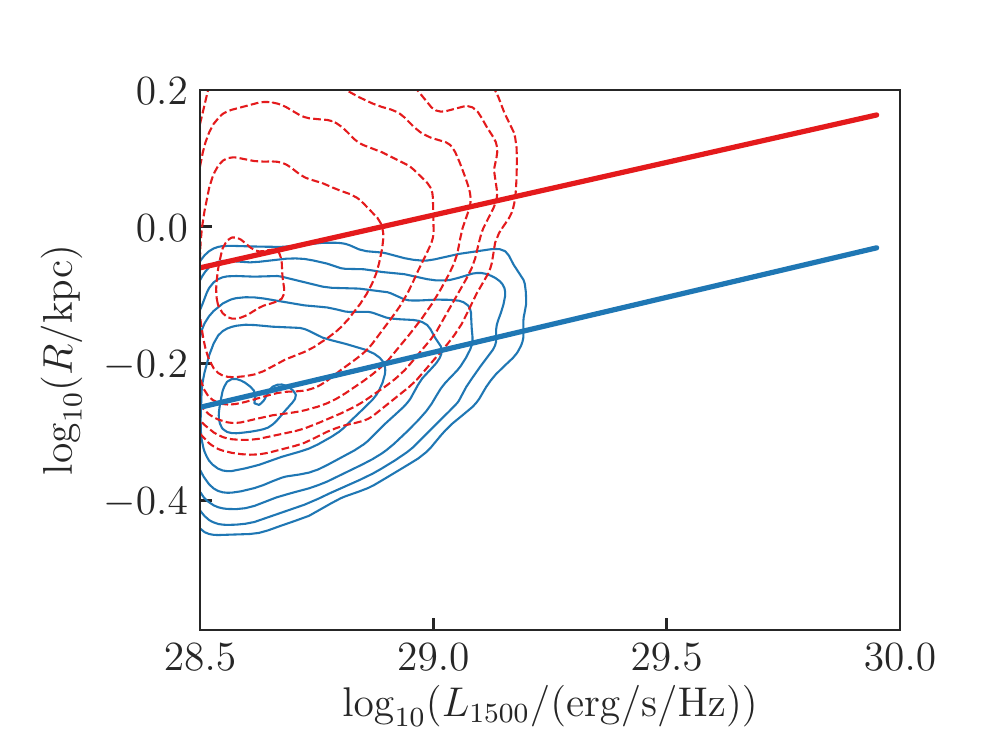}
\caption{Left: the $z=8$ size--mass relation and Right: the FUV size--luminosity relation. 
Shown are the relations for the galaxy half-light radius measured using circular aperture photometry, the half-mass radius inside
$R_{200}$ based on the 3D distribution of star particles, and the
effective radius. 
The half-mass radius is only included in the left panel, as this traces the intrinsic star particle 3D distribution in \BlueTidesns, with no equivalent luminosity calculation possible.
The half-light and effective radii are calculated from the mass maps in the left panel, and the dust-attenuated FUV (1500\AA) maps in the right panel. Lines show the best-fitting relation, and contours depict regions containing 10\%, 55\%, 77\%, 88\%, 94\%, and 97\% of galaxies in the corresponding samples.
}
\label{fig:AppSizeMeasures_SizeLuminosity}
\end{center}
\end{figure*}

In Figure \ref{fig:AppSizeMeasures_SizeLuminosity} we show the size--mass and size--FUV luminosity relations obtained using this half-light radius, compared with the effective radius. We find that the best-fitting size--mass relation is up to $\sim0.15$ dex larger for the half-light radius than the effective radius, with a slope that decreases more steeply to higher stellar masses.
The best-fitting size--luminosity relation is $\sim0.2$ dex larger for the half-light radius than the effective radius, with an offset roughly constant with luminosity.

Measuring larger sizes using the half-light radius curve-of-growth method is consistent with expectations. 
In the effective radius calculation, the effective area is equivalent to the minimum number of pixels in which half of the luminosity in the image is contained. These pixels \textit{do not} need to be contiguous. The effective radius is calculated from this area, by assuming the area is a circle, i.e. these brightest pixels are distributed as compactly as possible.
On the other hand, the half-mass radius is the radius of the smallest circle which contains half of the luminosity in the image. As these galaxies can be more extended along one direction, irregular, or clumpy, this circle will be larger than the circle in the effective radius calculation, which by definition assumes the most compact distribution possible.

The results presented throughout this work are qualitatively unchanged if the half-light radius is used instead of the effective radius, while the quantitative results will be altered. As the effective radius is a more robust measure for high-redshift galaxies, we opt to use this throughout.

In \citet{Marshall2020}, we investigated the intrinsic sizes of galaxies based on the 3D distribution of star particles. For this, we calculate the half-mass radius inside the galaxy $R_{200}$, the radius containing 200 times the critical stellar mass density, $R_{0.5}$.
This radius is also compared to the effective radius from the stellar mass map in Figure \ref{fig:AppRadiusComparison}. 
The majority of galaxies show similar effective radii and $R_{0.5}$, although there is large scatter.
In Figure \ref{fig:AppSizeMeasures_SizeLuminosity} we show the size--mass relation obtained using this half-mass radius $R_{0.5}$, compared with that of the effective radius. We find that the best-fitting size--mass relation for $R_{0.5}$ is $\sim0.15$ dex larger than that for the effective radius, with a shallower slope.
This radius cannot be used for analysing the size--luminosity relation, for example, as it is calculated from the stellar particle distribution, and does not consider the dust distribution, or mass-to-light conversion. Thus, it is not used throughout the rest of this work, with the effective radii more applicable and relevant to observational analyses.

\section{Quasar Host Galaxy Images}
\begin{figure*}
\begin{center}
\includegraphics[scale=0.8]{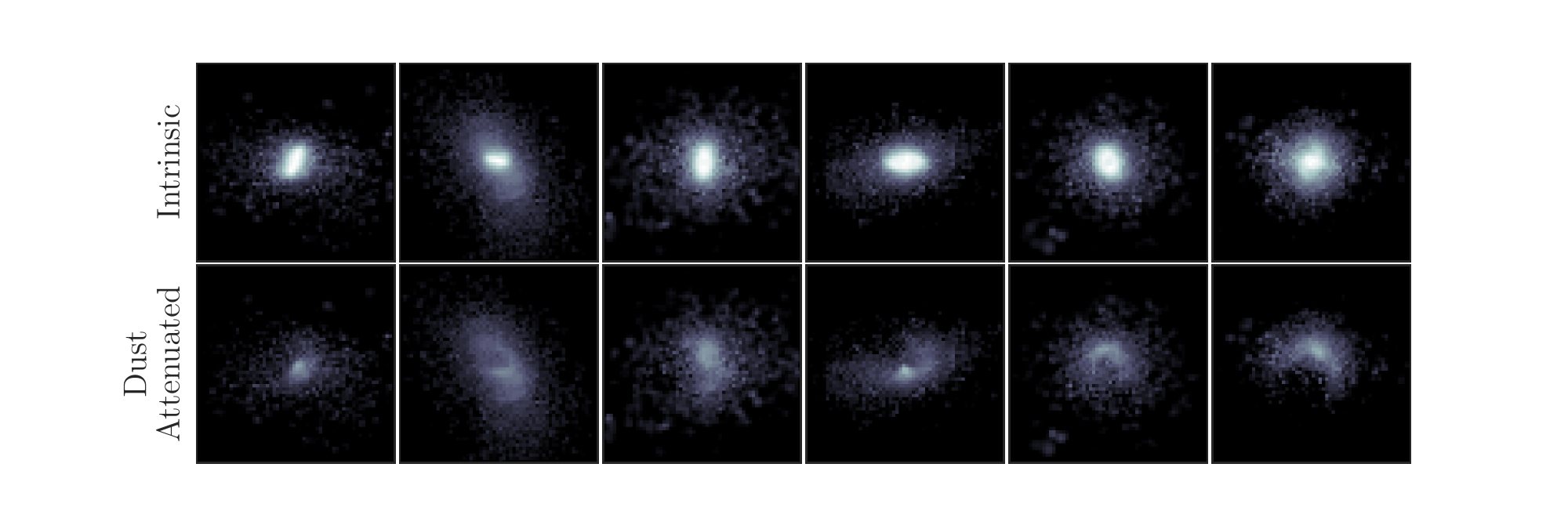}
\vspace{-0.5cm}

\includegraphics[scale=0.8]{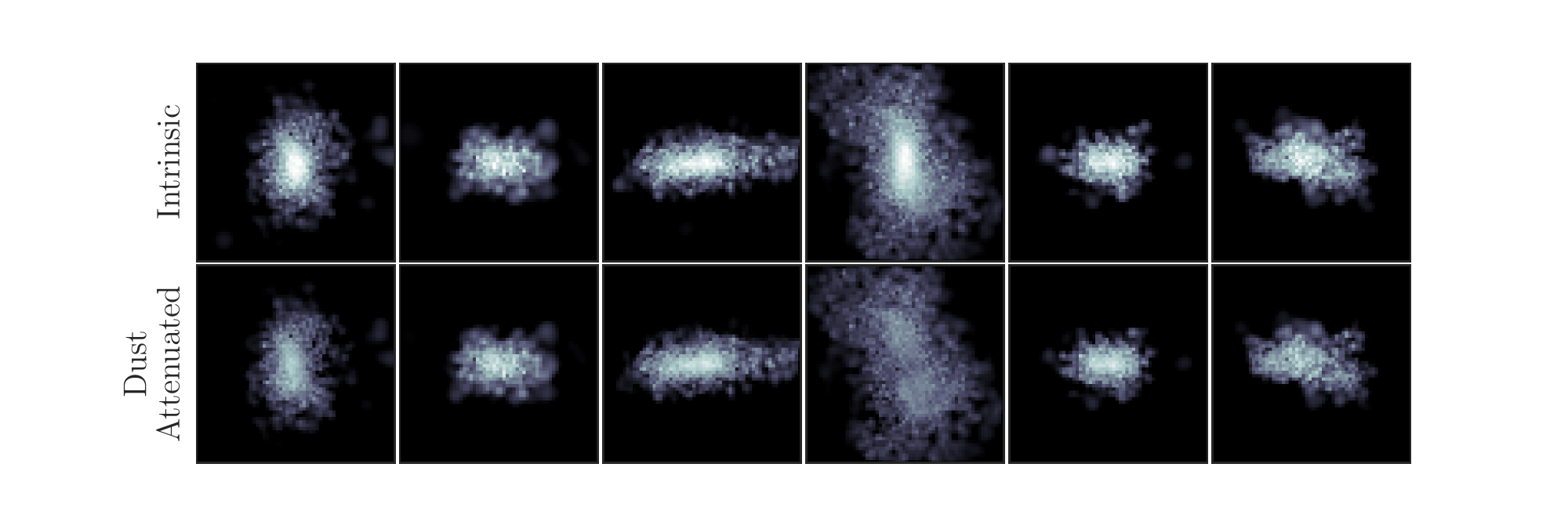}
\vspace{-0.3cm}
\caption{Images of six $z=7$ quasar host galaxies (top panels) and matched galaxies (bottom panels) in the V-band. The images show a $6\times6$ kpc FOV. The colour scale for each galaxy is the same for the pair of intrinsic and dust attenuated images, ranging from the maximum intrinsic pixel flux to 5\% of the maximum dust-attenuated pixel flux.
The quasar hosts have (left to right) stellar masses $M_\ast=4.22$, 7.08, 3.54, 6.53, 3.38 and 3.14 $\times10^{10}M_\odot$, and FUV luminosities $L_{1500}=9.78$, 43.0, 9.76, 15.1, 7.92, and 10.3 $\times10^{28}$ erg/s/Hz. The galaxies have (left to right) stellar masses $M_\ast=2.94$, 3.77, 2.20, 3.35, 1.81, and 1.67 $\times10^{10}M_\odot$, and FUV luminosities $L_{1500}= 9.62$, 13.5, 9.27, 20.2, 7.87, and 10.6 $\times10^{28}$ erg/s/Hz. 
The quasars have intrinsic V-band effective radii of (left-to-right) 2.19, 1.69, 2.46, 2.76, 2.59, and 2.82 kpc, and dust-attenuated V-band effective radii of 6.56, 7.09, 5.94, 6.72, 6.02, and 5.26 kpc.
The matched galaxies have intrinsic V-band effective radii of (left-to-right) 3.87, 4.55, 4.89, 4.51, 4.22 and 4.89 kpc, and dust-attenuated V-band effective radii of 5.44, 4.95, 5.44, 10.01, 4.37, and 5.59 kpc.}
\label{fig:QuasarImages}
\end{center}
\end{figure*}

Here in Figure \ref{fig:QuasarImages} we show V-band images of six of the bright quasar host galaxies at $z=7$, both with and without dust attenuation. For comparison, we also select six non-quasar host galaxies in the simulation with similar stellar masses and luminosities, and show equivalent V-band images both with and without dust attenuation. 

\bsp	
\label{lastpage}
\end{document}